\documentclass[]{interact}

\usepackage{epstopdf}
\usepackage[caption=false]{subfig}
\usepackage{multirow}
\usepackage{float}
\usepackage{pdflscape}
\usepackage{rotating}
\usepackage{graphicx} 
\usepackage[authoryear]{natbib}

\bibpunct{(}{)}{,}{a}{,}{,}

\renewcommand\bibfont{\fontsize{10}{12}\selectfont}
\usepackage{subcaption}
\theoremstyle{plain}

\theoremstyle{definition}

\theoremstyle{remark}

\usepackage{mathtools}

\usepackage{hyperref} 

\usepackage{algorithm}
\usepackage{algpseudocode}
\usepackage{adjustbox}
\usepackage[left]{lineno}

\newcommand{\answer}[1]{#1}
\usepackage{xcolor}
\usepackage{amsthm}

\begin{document}


\title{Modeling zero-inflated precipitation extremes}

\author{
\name{Aamar Abbas\textsuperscript{a} Touqeer Ahmad \textsuperscript{b}\thanks{CONTACT Touqeer Ahmad. Email: touqeer.ahmad@univ-orleans.fr; touqeer.ahmad8960@gmail.com} and Ishfaq Ahmad\textsuperscript{a}}
\affil{\textsuperscript{a}Department of Mathematics and Statistics, International Islamic University, Islamabad, Pakistan, and\\ \textsuperscript{b}Institute Denis Poisson, University of Orl\'eans, France.}
}

\maketitle

\begin{abstract}
Accurate modeling of daily rainfall, encompassing both dry and wet days as well as extreme precipitation events, is critical for robust hydrological and climatological analyses. This study proposes a zero-inflated extended generalized Pareto distribution model that unifies the modeling of dry days, low, moderate, and extreme rainfall within a single framework. Unlike traditional approaches that rely on prespecified threshold selection to
identify extremes, our proposed model captures tail behavior intrinsically through a tail index that aligns
with the generalized Pareto distribution.  The model also accommodates covariate effects via generalized additive modeling, allowing for the representation of complex climatic variability. The current implementation is limited to a univariate setting, modeling daily rainfall independently of covariates. 
Model estimation is carried out using both maximum likelihood and Bayesian approaches. Simulation studies and empirical applications demonstrate the model’s flexibility in capturing zero inflation and heavy-tailed behavior characteristics of daily rainfall distributions.
\end{abstract}


\section{Introduction}\label{sec:1}

Data characterized by excessive zeros and heavy-tailed distributions are prevalent in numerous fields, presenting unique challenges for statistical modeling. For instance, daily precipitation levels often exhibit a high frequency of zero values (days with no rainfall) along with a heavy-tailed distribution for positive rainfall amounts \citep{weglarczyk2005three}. Similarly, in the context of car insurance, the annual number of claims per client frequently includes a large proportion of policyholders with zero claims, whereas the remaining claims follow a skewed, heavy-tailed distribution \citep{christmann2004approach}. Another example can be found in healthcare, where the length of an overnight hospital stay per patient often includes a significant clump at zero (patients discharged on the same day) and a right-skewed distribution for longer stays \citep{chen2007hospital}. In radio audience measurements, daily listening minutes for a given station are highly skewed and zero-inflated because of non-listeners  \citep{Couturier2010}. These examples underscore the need for robust statistical models that can effectively handle datasets with excess zeros and heavy-tailed distributions.

Zero-inflation is a particularly common problem in daily and hourly precipitation data. For instance, rainfall measurements often include days with no precipitation, which results in a high frequency of zeros. In contrast, the positive values (rainy days) exhibited a heavy-tailed distribution owing to occasional extreme rainfall events. In various studies, researchers have attempted to model only the extreme right tail using the peak over threshold (POT) method, which models exceedances beyond a high threshold. When the threshold is sufficiently high, the distribution of exceedances often approximates the Generalized Pareto Distribution (GPD). A significant limitation of GPD is that it can only model observations that exceed a specified high threshold, leaving uncertainty about the behavior of the remaining observations below this threshold. Various attempts have been made to model the entire range of data using a mixture of models. \answer{\citet{kedem1990estimation} highlighted that rainfall is best represented by a mixed distribution, with a discrete spike at zero for dry periods and a continuous distribution (lognormal or gamma) for wet periods.} \citet{frigessi2002dynamic} proposed a dynamic mixture model for unsupervised tail estimation. Using a weight function, the model combined the light- and heavy-tailed densities of the GPD. \citet{carreau2009hybrid}  proposed a hybrid Pareto distribution by extending GPD that can be used in a mixture model to model the entire real axis.  \citet{scarrott2012review} offers a comprehensive overview of mixture modeling and its applications by introducing the mixture form of two models below and above the specified threshold. \citet{papastathopoulos2013extended} introduces a new extension of GPD that models low-threshold exceedances and retains the same tail behavior as GPD. A drawback of these approaches is the need to select an appropriate threshold and an increased number of parameters, which complicates the inference. Following \citet{papastathopoulos2013extended}, \citet{naveau2016modeling} proposed extensions of GPD that are sufficiently flexible to model the entire range of positive rainfall intensities without a threshold selection step. They employed censored likelihood methods to handle low precipitation values effectively. 
\citet{stein2021parametric} introduced both tails of the flexible distribution to model the tails of the data distribution by the tail index of GDP. \answer{
\citet{wilson2022beyond} proposed a deep extreme mixture model that offers a unified framework for better capturing moderate events, including zero and extreme events, while allowing flexible thresholding for spatio-temporal forecasting of precipitation. \citet{haruna2025joint} applied a non-stationary extended GPD to model long-term trends in both the bulk and the extremes of daily precipitation distributions in Switzerland, while dry events were treated separately in their modeling strategy.} Overall, the issue of zero inflation in precipitation data has been largely overlooked despite its critical impact on statistical modeling.

To address the challenge of modeling zero inflation and heavy tails together, this study proposes a Zero-inflated Extended Generalized Pareto Distribution (ZIEGPD) for modeling zero-inflated precipitation data. ZIEGPD has a tail index parameter similar to that of GPD. Therefore, the proposed model is well-suited for modeling heavy-tailed data and excessive zeros, and the distribution bypasses the threshold choice step, making it an ideal candidate for describing dry and wet events simultaneously. By extending the \answer{EGPD} to a zero-inflated framework, our model can simultaneously estimate the probability of zero precipitation and the distribution of positive rainfall amounts, providing a comprehensive approach for analyzing zero-inflated heavy-tailed data. Several constructions have been proposed in the literature to handle zero inflation in data. The most commonly applied method for handling zero inflation was introduced by \citet{lambert1992zero} by mixing a two-part model specifically for continuous data zero-inflated models to account for excess zeros. \citet{ahmad2024extended, ahmad2024new} proposed a flexible model for zero-inflated heavy-tailed discrete data.\answer{\citet{dzupire2018poisson}showed that modeling rainfall occurrence and intensity separately tends to underestimate the variability inherent in rainfall data, and they proposed a Poisson–Gamma model that jointly accounts for both zero and non-zero rainfall. Ensemble precipitation forecasts are often unreliable due to zero-inflation, heteroscedasticity, and rare extreme events. To address limitations of separate modeling of occurrence and intensity, \cite{scheuerer2015statistical} proposed a censored, shifted gamma distribution (CSGD) model that jointly models zeros and precipitation amounts, improving probabilistic forecast.}

Section~\ref {method} describes the proposed model as a natural extension of the aforementioned models for zero-inflated heavy-tailed continuous data. The proposed ZIEGPD is quite flexible at the lower tail of the distribution, while correctly modeling the zero proportion and tail of the distribution. For estimation purposes, we introduce maximum likelihood and Bayesian estimation paradigms.

The remainder of this paper is organized as follows. Section~\ref{method} provides the background of zero-inflated mixture models and introduces the proposed model. Section~\ref{inference} discusses the inferential procedures. Detailed simulation results are provided in Section~\ref {SimulationStudy}. The real-data application findings are explained in Section~\ref{realapp}. Section~\ref{concl} concludes the paper.

\section{Modeling framework}\label{method}
Under the umbrella of extreme value theory, the GPD is considered appropriate for modeling extreme observations over a high threshold. The GPD is defined by its Cumulative Distribution Function (CDF) as
\begin{equation} \label{CDFGPD}
	\mathcal{H}(z;\sigma,\xi) =
	\begin{cases} 
		1-\left(1+\xi  z/\sigma\right)_{+}^{-1/\xi} & \xi\neq 0 \\
		1- \exp{(-{z}/{\sigma})} &  \xi = 0
	\end{cases}
\end{equation}
where $(a)_{+}= \max(a,0)$, $\sigma>0$ and $-\infty<\xi<+\infty$ represent the scale and shape parameters of the distribution, respectively. 

The shape parameter $\xi$ of the GPD characterizes the tail behavior. If $\xi<0$, the upper tail is bounded. If $\xi=0$, the distribution approaches an exponential form with all moments finite. If $\xi>0$, the upper tail is unbounded, but higher moments ultimately become infinite.
 
In GPD approximation, selecting the optimal threshold is critical but challenging. A low threshold can introduce bias by including non-extreme values, which distorts tail estimation. Conversely, a high threshold limits the number of exceedances, leading to higher variance and unreliable estimates due to the small sample size.

In contrast, some continuous extreme value models aim to avoid the threshold selection by modeling the entire range of data. For instance, \citet{papastathopoulos2013extended} extended the GPD model by adding an additional shape parameter that does not alter the tail behaviour. This modification enhances the stability of threshold selection, enabling the use of a lower threshold, which improves the modeling of both the upper tail and the bulk of the distribution. \citet{naveau2016modeling} further outlined two critical conditions that the model must satisfy to remain consistent with the requirements of EVT.

For the development of the models proposed in this study, we \answer{follow} the way outlined by \citet{naveau2016modeling}, which uses the idea of integral transformation to simulate random draws from a GPD. Specifically, their method involves generating GPD random variables by applying the inverse of the GPD CDF in \eqref{CDFGPD}, denoted by $\mathcal{H}_{\sigma,\xi}^{-1}(U)$, where $U \sim \mathcal{U}(0,1)$ is a random variable uniformly distributed on the interval $(0,1)$. We can construct a family of models through the idea of inverse transformation for random variables
\begin{equation}\label{eq:3}
Z=\mathcal{H}_{\sigma,\xi}^{-1}\left(\mathcal{W}^{-1}(U) \right),
	\end{equation}
where $\mathcal{W}$ is a CDF on $[0,1]$ and $U\sim\mathcal{U}(0,1)$. The CDF of $Z$ can be written as $\mathcal{W}(\mathcal{H}(z;\sigma,\xi))$.
The key challenge is to identify a function \(\mathcal{W}\) that retains the upper tail behavior characterized by the shape parameter \(\xi\), while also regulating the lower tail behavior. \citet{naveau2016modeling} proposed conditions that a valid \(\mathcal{W}\) function must satisfy. Specifically, the tail of \(\mathcal{W}\), denoted as \(\Bar{\mathcal{W}} = 1 - \mathcal{W}\), must meet the following requirements:
\begin{eqnarray}
    \lim_{u \to 0} \frac{\Bar{\mathcal{W}}(1-u)}{u} &=& a, \quad \text{for some constant } a > 0 \text{ (upper tail behavior)}, \label{eq:cond1} \\
    \lim_{u \to 0} \frac{\mathcal{W}(u)}{u^\kappa} &=& c, \quad \text{for some constant } c > 0 \text{ (lower tail behavior)}. \label{eq:cond2}
\end{eqnarray}
 For CDF of $\mathcal{W}$, \answer{we consider three families proposed by \citet{naveau2016modeling}, denoted by $\mathcal{W}(\cdot,\Psi)$,}that satisfy the conditions outlined in \eqref{eq:cond1} and \eqref{eq:cond2}\\
	\textbf{\textit{Model 1 ($M_1$)}}.  \label{c1} 
	$\mathcal{W}(u;\Psi)=u^{\kappa}$, $\Psi=\kappa >0$;\\
	\textbf{\textit{Model 2 ($M_2$)}}. \label{c3}
	$\mathcal{W}(u;\Psi)=1-B_{\delta}\{(1-u)^{\delta}\}$, $\Psi=\delta>0$ where $B_{\delta}$ is the CDF of a Beta random variable with parameters $1/\delta$ and $2$, that is:
	$$B_{\delta}(u)=\frac{1+\delta}{\delta} u^{1/\delta} \Big( 1-\frac{u}{1+\delta} \Big)$$\\
	\textbf{\textit{Model 3 ($M_3$) }}. \label{c4}
	$\mathcal{W}(u;\Psi)=[1-B_{\delta}\{(1-u)^{\delta}\}]^{\kappa/2}$, $\Psi=(\delta, \kappa)$ with $\delta>0$  and $\kappa>0$.

The CDF \(\mathcal{W}(\mathcal{H}(z; \sigma, \xi))\), referred to as the CDF of Extended GPD (EGPD), is valid only for modeling non-zero observations or low threshold exceedances. \answer{For nonzero precipitation data, a censored likelihood approach is recommended to improve estimation, with low nonzero values rounded (e.g., to 0.1 for hourly data) to reduce discretization effects \citep{naveau2016modeling}.}

In numerous practical scenarios, datasets frequently exhibit an excessive number of zeros. These excessive zeros complicate precise statistical analysis, as standard statistical models are typically not designed to handle such cases. \answer{Thus, it is essential to employ modeling approaches that reduce bias and properly capture the data structure. 
One effective approach is to use mixture models that jointly account for zero inflation and non-zero observations. }

\subsection{Zero inflated mixture model (ZIMM)}
A robust method for modeling data characterized by excessive zeros uses mixture models, specifically the Zero-Inflated Distribution (ZID). \cite{lambert1992zero} can be seen as one of the initial references to modeling zero-inflated data. The ZID models require a specific distribution to model the non-zero values and a degenerate distribution concentrated at zero \citep{rodrigues2003bayesian}. The mixture distribution for integer variable with zero-inlation is defined as
\begin{equation}\label{eq:ZIDEGPD}
		\Pr(X=x|\Theta=(\theta, \pi))=\left\{
		\begin{array}{ll}
			\pi +(1-\pi)g(x|\theta) & if \hspace{0.1cm} x=0 \\
			(1-\pi)g(x|\theta) & if \hspace{0.1cm} x \in \mathbb{N}, 
		\end{array}
		\right.
	\end{equation}
 where $0\leq \pi \leq 1$ and $\theta$ is vector of parameters. The $\pi$ is the proportion of observations that inflates the data distribution. We denote \(\pi\) as the weight assigned to zero values and \((1 - \pi)\) as the weight for non-zero observations. The mixture distribution can be expressed as follows 
 \begin{equation}\label{mixture_m}
     \Pr(x|\theta, \pi)= \pi \cdot I_0(x) + (1 - \pi) \cdot \Pr(y|\theta),
 \end{equation}
 where $I_0(x)$ is a degenerate distribution concentrated at the zero and $\Pr(y|\theta)$ is a probability function
that fits the data. The likelihood function can then be simplified by employing a procedure that utilizes extended data with latent variables, see for instance, \citet{rodrigues2003bayesian}.

\subsection{Zero inflated extended generalized Pareto (ZIEGP) distribution}
The ZIEGP distribution is proposed to address the challenges of modeling data that exhibits extreme values and a high frequency of zeros, which is commonly seen in precipitation data when there is no rain. As mentioned above, the EGPD is typically valid for modeling non-zero observations, while the proposed zero-inflated version serves as an effective alternative for capturing excessive zeros in data. To construct ZIEGPD, we employ a mixture approach that integrates the ZIMM with the EGPD model. This strategy enables us to adeptly model datasets exhibiting zero inflation and extreme value properties, thereby providing a more versatile and precise analytical framework for relevant use.

The density function of ZIEGP distribution is derived by replacing $\Pr(y|\theta)$ in \eqref{mixture_m} with the density function of EGPD. Mathematically, the density function is defined as
\begin{equation}\label{den-ziegpd}
    h(z|\pi, \sigma, \xi; \Psi)=\pi \cdot I_0(z) + (1 - \pi) \cdot \frac{d}{dz}  \mathcal{W}(\mathcal{F}(z; \sigma, \xi); \Psi), 
\end{equation}
where $z\geq0$, $0\leq \pi \leq 1$, $I_0(x)$ is the indicator function and $\mathcal{W}$ denotes the families which we define previous and $\Psi$ is parameter vector of $\mathcal{W}$. For every $\mathcal{W}$ family, the mathematical derivation of \eqref{den-ziegpd} can be found in the Appendix.

By integrating the density function of ZIEGPD given in \eqref{den-ziegpd}, we found the CDF of ZIEGPD as
\begin{equation}
     H(z|\pi, \sigma, \xi; \Psi)=  \pi \cdot I_0(z) + (1 - \pi) \mathcal{W}(\mathcal{F}(z; \sigma, \xi); \Psi)
\end{equation}
where $z\geq0$, $0\leq \pi \leq 1$. \answer{The quantile function is defined as}

 \begin{equation}\label{eq:quantile}
 	q_{p^*}=H^{-1}(p^*)
 \left\{
 	\begin{array}{ll}
 		\frac{\sigma}{\xi}\left\{ \left(1- \mathcal{W}^{-1}(p^*) \right)^{-\xi}-1\right\}, & \mbox{if } \xi>0\\
   \\
 		-\sigma \log \left(1- \mathcal{W}^{-1}(p^*) \right),& \mbox{if } \xi=0,\\
 	\end{array}
 \right.
 \end{equation} 
 where $p^*=\frac{p-\pi}{1-\pi}\in (0, 1)$. In EVT, the $p^{th}$ helps to estimate the return level $R_t$, which is the value expected to be exceeded once every 
$t$ time periods. This is expressed by the quantile $p=1-\frac{1}{t}.$

\begin{figure}{}
\centering
\subfloat[]{%
\resizebox*{8cm}{!}{\includegraphics{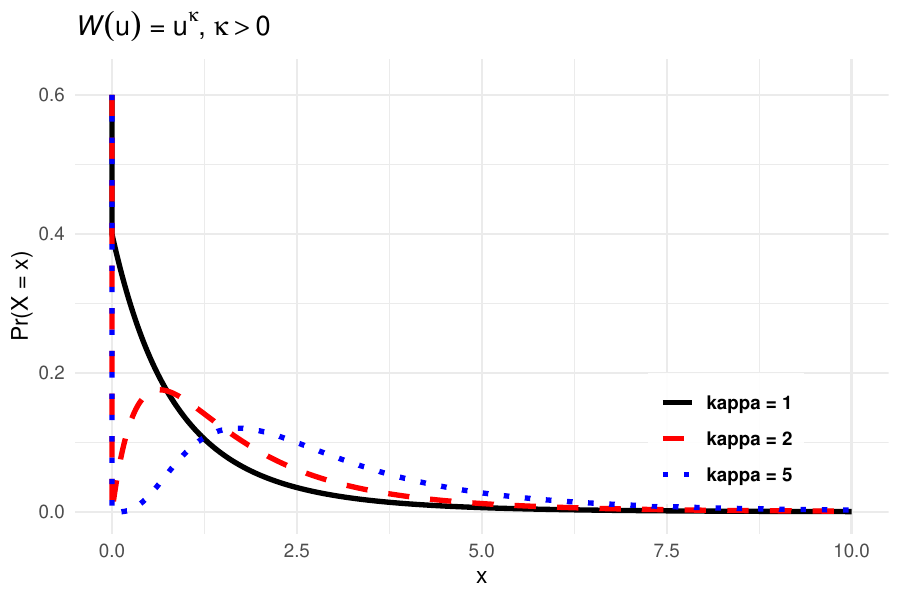}}}
\subfloat[]{%
\resizebox*{8cm}{!}{\includegraphics{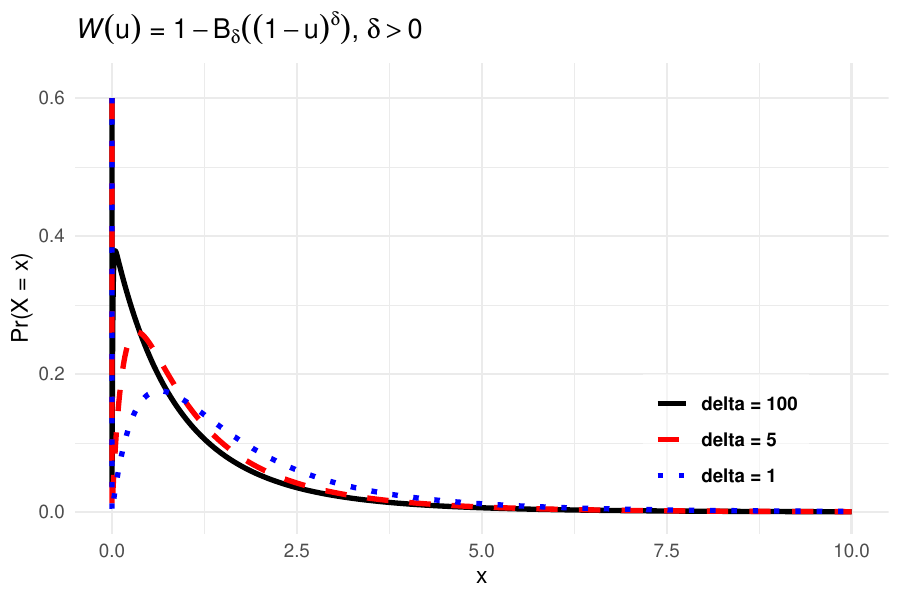}}}\\
\subfloat[]{%
\resizebox*{8cm}{!}{\includegraphics{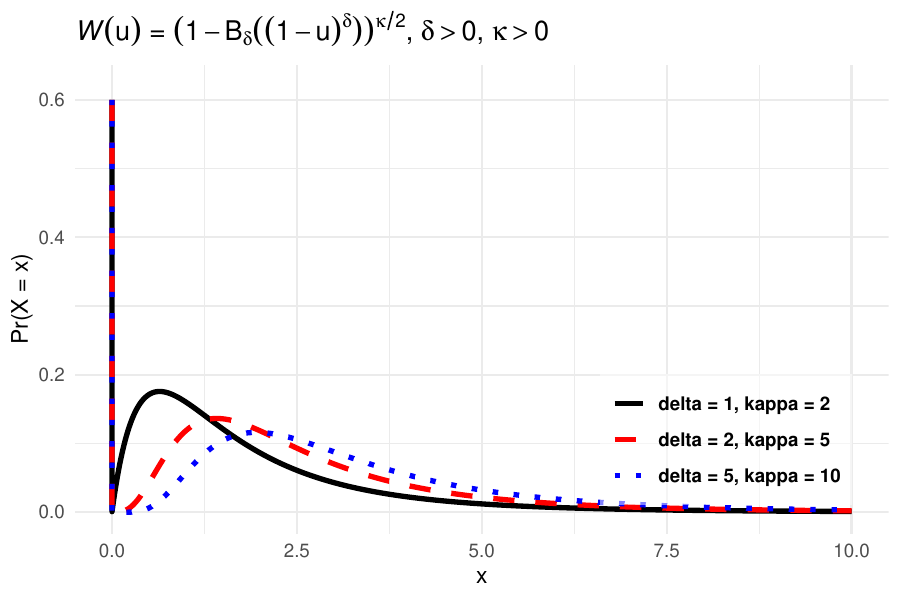}}}
 \caption{Flexible density function behaviors of model proposed in \eqref{den-ziegpd} by fixing parameters $(\pi=0.6, \sigma=1, \xi=0.5)$ and varying parameter $\kappa=1, 2$ and $5$ for model $\mathcal{W}(u;\Psi)=u^{\kappa}$, $\Psi=\kappa >0$ (top-left), varying parameter $\delta=100, 5$ and $1$ for model $\mathcal{W}(u;\Psi)=1-B_{\delta}\{(1-u)^{\delta}\}$, $\Psi=\delta>0$ (top-right) and varying parameters $(\delta, \kappa)=(1, 2), (2, 5) $ and $(5, 10)$ for model $\mathcal{W}(u;\Psi)=[1-B_{\delta}\{(1-u)^{\delta}\}]^{\kappa/2}$, $\Psi=(\delta, \kappa)$ with $\delta>0$  and $\kappa>0$ (bottom).  }
    \label{fig:density} 
\end{figure}

The expression \eqref{den-ziegpd} with incorporation of $\mathcal{W}(u;\Psi)=u^{\kappa}$, $\Psi=\kappa >0$ leads to the density function of ZIEGPD  with parameters $(\pi, \kappa, \sigma,$ and $\xi)$, where $\pi$ is the parameter for proportion of zeros, $\kappa$ is the shape parameter for the lower tail of the distribution and $\sigma$ and $\xi$ are usual scale and shape parameters. Figure~\ref{fig:density} (top-left) illustrates the density behavior with fixed values for $(\pi=0.60, \sigma = 1$ and $\xi = 0.5)$ while varying the lower tail behavior by changing $\kappa$ to 1, 2, and 5. When $\kappa$ is set to 1, the model reduces to the \answer{GPD} with the addition of zero inflation. Increasing $\kappa$ introduces greater flexibility in the lower tail behavior along with same amount of zero inflation and without compromising the characteristics of the upper tail, as evidenced in Figure~\ref{fig:density} (top-left) when $\kappa$ is set to 5.

The expression \eqref{den-ziegpd} with combination of  $\mathcal{W}(u;\Psi)=1-B_{\delta}\{(1-u)^{\delta}\}$, $\Psi=\delta>0$ leads to another density function of ZIEGPD  with parameters $(\pi, \delta, \sigma,$ and $\xi)$, where $\delta$ is characterizes the central portion of the distribution and enhancing flexibility
in that region. In contrast, other parameters have the same role as we discussed above. With varying $\delta$ values, Figure~\ref{fig:density} (top-right) shows different behaviors. For large values of $\delta$, the distribution behaves similarly to GPD with an additional number of zeros at the origin. \answer{Decreasing $\delta$ values only show flexibility at the central part of the distribution without affecting the upper tail behavior and keeping the same zero inflation at the origin, while the lower tail remains difficult to estimate reliably from the data (see Figure~\ref{fig:density} (top-right)). This drawback has also been noted by \citet{naveau2016modeling}.}  

To overcome the drawback seen in $\mathcal{W}(u;\Psi)=1-B_{\delta}\{(1-u)^{\delta}\}$, $\Psi=\delta>0$, we introduce flexibility at \answer{lower tail} of distribution by additional parameter $\kappa$, the expression \eqref{den-ziegpd} with combination of  $\mathcal{W}(u;\Psi)=[1-B_{\delta}\{(1-u)^{\delta}\}]^{\kappa/2}$, $\Psi=(\delta, \kappa)$ with $\delta>0$  and $\kappa>0$ leads to another flexible density function of ZIEGPD  with parameters $(\pi, \delta, \kappa, \sigma,$ and $\xi)$, where $\delta$ is characterizes the central part of the distribution and $\kappa$ creates flexibility at \answer{lower tail} of the distribution. With varying $\kappa$ values, Figure~\ref{fig:density} (bottom) shows flexibility at the lower part of the distribution, even when varying $\delta$ to high values. For instance, at $(\delta=5, \kappa=10 )$, the clear flexibility at the lower part can be observed with the inflation of zeros in the model.

\section{Inference for ZIEGP model}\label{inference}
This section introduces the methods for estimating the ZIEGP model. For the purpose of obtaining better results, even in the case of nonstationary data. \\
The joint likelihood function of the ZIEGP model is derived as follows
\begin{equation}\label{lik}
    L(\pi, \sigma, \xi, \Psi|Z)= \prod_{i=1}^n\Bigg[\pi \cdot I_0(z_i) + (1 - \pi) \cdot \frac{d}{dz}  \mathcal{W}(\mathcal{F}(z_i; \sigma, \xi); \Psi)\bigg]. 
\end{equation}
After solving the derivative involved in \eqref{lik} for each $\mathcal{W}$ choice, the expressions for the density functions are provided in Appendix~A. The log likelihood is defined as 
\begin{equation}\label{loglik}
    l(\pi, \sigma, \xi, \Psi|Z)= \sum_{i=1}^{n}I_0(z_i) \log \pi   + \sum_{i=1}^{n}\log \bigg[ (1 - \pi) \cdot \frac{d}{dz}  \mathcal{W}(\mathcal{F}(z_i; \sigma, \xi); \Psi)\bigg]. 
\end{equation}
In case of non-stationary data, we also developed the non-stationary version of the proposed models along with classical and Bayesian paradigms, see Appendix~B.
From an application perspective, the proposed models do not consider incorporating covariates; however, the developed methodology remains adaptable for their inclusion if desired. For the classical approach, we implement the proposed models using our own developed R code and incorporate it as a \texttt{custom families} in the \texttt{evgam} \citep{evgam-youngman} R package. For Bayesian estimation, we introduce the models as an add-on in the \texttt{bamlss} R package \citep{bamlss}. For all parametric, linear effects, we follow \citet{Klein2015} and assume a flat, noninformative prior for the intercept, representing a limiting case of a multivariate Gaussian prior with high dispersion. This approach can provide regularization for high-dimensional parameter vectors. 

\section{Simulation Study}\label{SimulationStudy}
This section presents a simulation study for our proposal in two stages. First, we perform a model-based simulation, generating data directly from our model to assess its accuracy in estimating the parameters. In the second stage, we generate data from an alternative model, such as zero-inflated GEV (ZIGEV), to evaluate the robustness of our model estimates when the data-generating process differs.  
\subsection{Model based simulation}
We simulated the data from our proposed model with different parameter configurations. We set $\sigma$ the scale parameter set to 1 in all configurations, the parameter for a proportion of zero varies (i.e., $\pi=0.2, 0.5$), and the upper tail index parameter also varies (i.e., $\xi=0.2, 0.3, 0.4$). Futher, the $\kappa$ parameter for model $(\mathcal{W}(u;\Psi) = u^{\kappa}, \quad \Psi = \kappa > 0)$ is set to $5$ and $10$, while $\delta$ parameter for model $(\mathcal{W}(u;\Psi) = 1 - B_{\delta}\{(1 - u)^{\delta}\}, \quad \Psi = \delta > 0)$ is chosen as $1$ and $5$. For model $(\mathcal{W}(u;\Psi) = \left[1 - B_{\delta}\{(1 - u)^{\delta}\}\right]^{\kappa/2}, \quad \Psi = (\delta, \kappa)>0 )$, the parameters $(\delta, \kappa)=(1,5), (5,10) $ are used. For each configuration, we estimated the model parameters across 
$10^4$
  replications with sample sizes $n=500, 1000$, employing both maximum likelihood estimation (MLE) and Bayesian methods.

\answer{The boxplots of estimated parameters of all three models through MLE and Bayesian are shown in Figure~\ref{fig:model-based-sim}, for sample size $n=1000$. Figure~\ref{fig:model-based-sim-Sup} in the supplement file reports the results for $n=500$. The top-left panel in both figures illustrates the results obtained by fitting Model $M_1$ using both classical and Bayesian paradigms.} Boxplots depict the estimation performance of the true parameters across all configurations. Both methods accurately estimate the true parameters, demonstrating robustness across scenarios. Furthermore, the proportion of zeros and bulk parameter $\kappa$ are consistently estimated even in cases where the values $\xi$ are relatively large, i.e., $\xi=0.4$.
\begin{figure}
\centering
\subfloat[$M_1$ with $n=1000$]{%
\resizebox*{7cm}{!}{\includegraphics{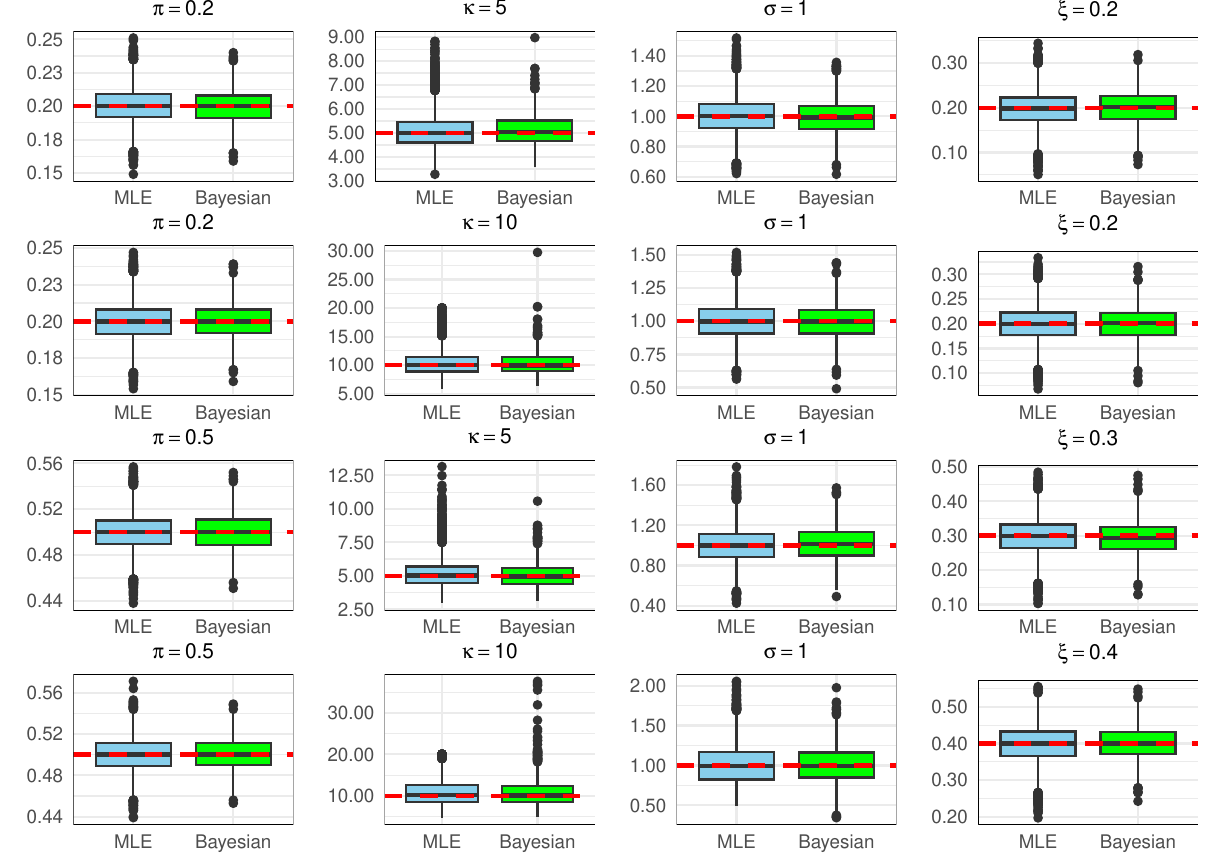}}}
\subfloat[$M_2$ with $n=1000$]{%
\resizebox*{7cm}{!}{\includegraphics{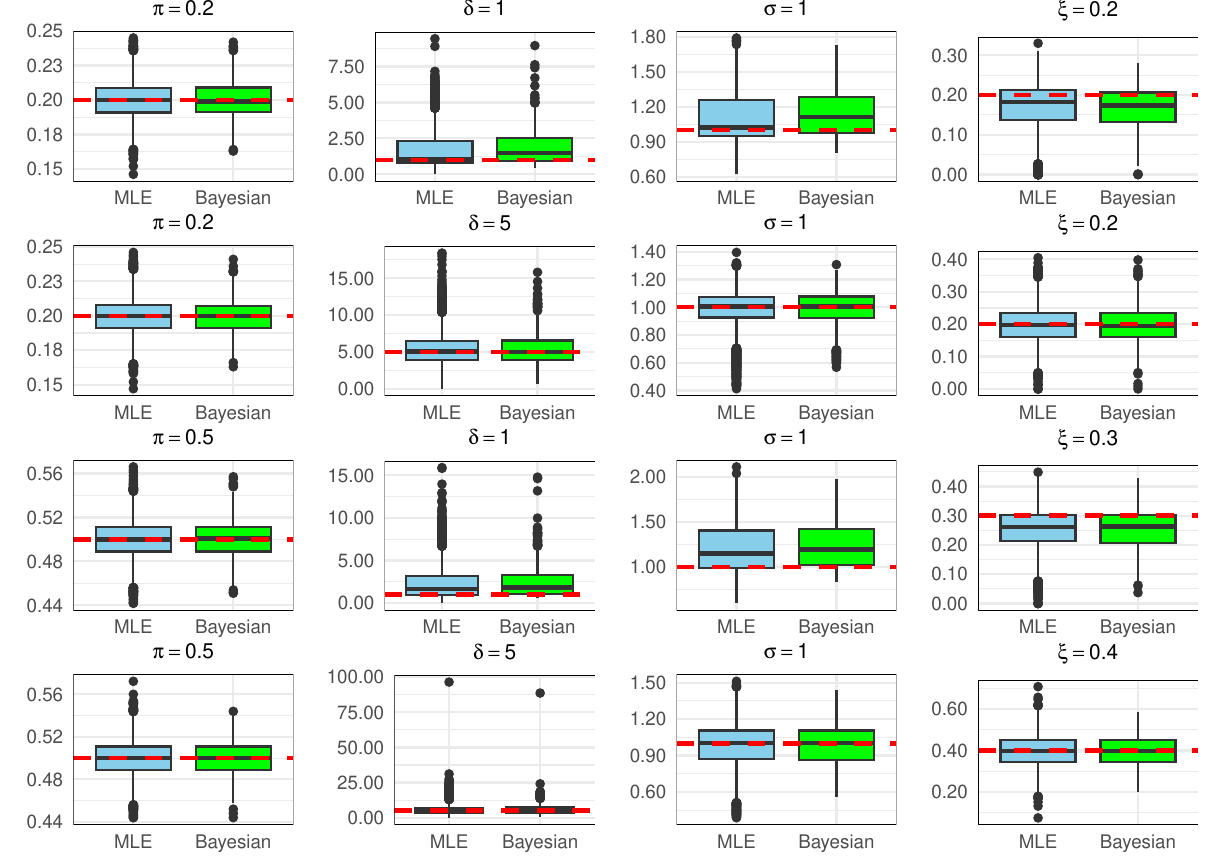}}}\\
\subfloat[$M_3$ with $n=1000$]{%
\resizebox*{7cm}{!}{\includegraphics{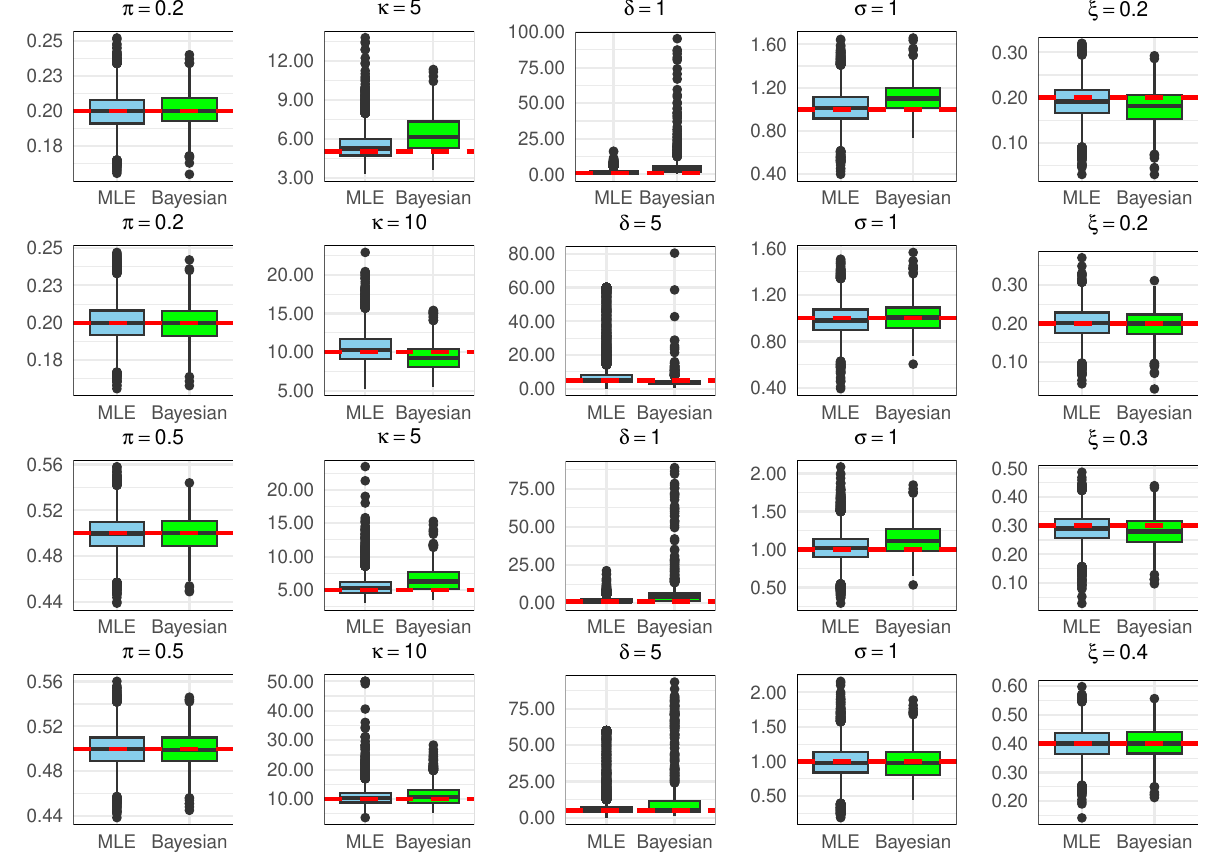}}}
\caption{\answer{Box plots showing the maximum likelihood estimates and Bayesian estimates of model parameters for each model type, based on sample sizes of $n = 1000$, with $10^4$ replications under varying parameter settings. 
}}
\label{fig:model-based-sim}
\end{figure}

\answer{The right and bottom panels in both figures present boxplots} of the estimated parameters obtained by fitting models $M_2$ and $M_3$, respectively. In both models, the parameter $\delta$, which governs the skewness, proves challenging to estimate accurately. The estimates of $\delta$ exhibit greater variability across configurations, regardless of whether the classical or Bayesian method is employed. This increased variation underscores the inherent difficulty in reliably estimating parameters related to skewness in these models. Such estimation difficulties of the skewness parameter of models were also noticed in literature by \citet{sartori2006bias, ribereau2016skew}.  
In the estimation of remaining parameters, the MLE performs better in small samples, while Bayesian estimates look more stable in large sample settings. Again, the parameter for the zero proportion is estimated correctly with both methods, regardless of the sample size choice.

To enhance the comparison between both estimation approaches, Table \ref{RMSE-M1} presents the root mean square errors (RMSEs) for both MLE and Bayesian estimates corresponding to $M_1$ when the sample size varies from $n=500$ to $n=1000$. The RMSEs results of models $M_2$ and $M_3$ are presented in Table~\ref{RMSE-M2} and Table~\ref{RMSE-M3}, respectively. From the RMSEs in the Tables, the MLE estimators show slightly better performance when sample size $n=500$, while the Bayesian method shows slight improvement when sample size increases. For instance, both estimators have similar performance with $\kappa \geq 5$ and $\xi=0.3, 0.4$, and Bayesian shows superiority when $\xi=0.2$ and $\pi=0.2$. In Model 2, with a larger $n$, the Bayesian method again shows better performance, apart from the larger variation in $\delta$. For model $M_3$, both estimators almost have similar performance, but MLE slightly outperforms in the case of a small sample, while Bayesian looks superior in a large sample setting when the true parameters are $\pi=0.2$ and $\xi=0.2$. In addition to the discussion of the results of models $M_2$ and $M_3$, the Bayesian method surprisingly gained higher variability in skewness parameter $\delta$ compared to MLE.    
\begin{table}[ht]
    \centering
    \caption{Root mean square errors for the parameters of $M_1$ estimated through classical and Bayesian when sample size $n=500, 1000$ used. }
    \begin{tabular}{ccccccccccccccccccc}
        \toprule
        \multicolumn{12}{c}{$\mathcal{W}(u;\Psi)=u^{\kappa}$, $\Psi=\kappa >0$ when $n=500$} \\ 
        \midrule
        & \multicolumn{2}{c}{RMSE} & & \multicolumn{2}{c}{RMSE} & & \multicolumn{2}{c}{RMSE} & & \multicolumn{2}{c}{RMSE} \\
        \cmidrule(lr){2-3} \cmidrule(lr){5-6} \cmidrule(lr){8-9} \cmidrule(lr){11-12}
        $\pi$ & $\text{MLE}$ & $\text{Bayes}$ & $\kappa$ & $\text{MLE}$ & $\text{Bayes}$ & $\sigma$ & $\text{MLE}$ & $\text{Bayes}$ & $\xi$ & $\text{MLE}$ & $\text{Bayes}$\\
        \midrule
        0.2   & 0.012 & 0.018 & 5 & 0.69 & 1.00 & 1 & 0.11 & 0.16 & 0.2 & 0.04 & 0.05 \\
        0.2  & 0.012 & 0.019 & 10 & 2.07 & 3.26 & 1 & 0.14 & 0.19 & 0.2 & 0.03 & 0.05 \\
        0.5   & 0.016 & 0.023 & 5 & 1.00 & 1.82 & 1 & 0.17 & 0.25 & 0.3 & 0.05 & 0.07 \\
        0.5  & 0.016 & 0.023 & 10 & 3.41 & 7.26 & 1 & 0.24 & 0.35 & 0.4 & 0.05 & 0.07 \\
        \midrule
        \multicolumn{12}{c}{$\mathcal{W}(u;\Psi)=u^{\kappa}$, $\Psi=\kappa >0$ when $n=1000$}\\
        \midrule
        0.2   & 0.013 & 0.012 & 5 & 0.68 & 0.68 & 1 & 0.12 & 0.12 & 0.2 & 0.04 & 0.04 \\
        0.2  & 0.013 & 0.012 & 10 & 2.02 & 1.99 & 1 & 0.14 & 0.13 & 0.2 & 0.03 & 0.03 \\
        0.5   & 0.016 & 0.016 & 5 & 1.02 & 0.96 & 1 & 0.17 & 0.17 & 0.3 & 0.05 & 0.05 \\
        0.5  & 0.016 & 0.016 & 10 & 3.56 & 3.96 & 1 & 0.25 & 0.24 & 0.4 & 0.05 & 0.05 \\
        \bottomrule
    \end{tabular}\label{RMSE-M1}
\end{table}

\subsection
{Tail stability with varying zero inflation}

To observe the right tail behavior with varying levels of zero inflation, we simulate $10^4$ heavy-tailed synthetic samples of size $n = 1000$ with different amounts of zeros from a zero-inflated generalized extreme value distribution \citep{quadros2020bayesian}. We use the following parameter settings: $\pi = 0.2, 0.4, 0.6, 0.8$, $\mu = 2$, $\sigma = 1$, and $\xi = 0.2$. Here, $\mu = 2$ is chosen to ensure that only nonnegative values are generated for the nonzero part of the samples. Additionally, we impose the condition that any negative observation generated in the samples be immediately replaced with the minimum nonzero positive value.
\begin{figure}
\centering
\subfloat[Model $M_1$]{%
\resizebox*{6cm}{!}{\includegraphics{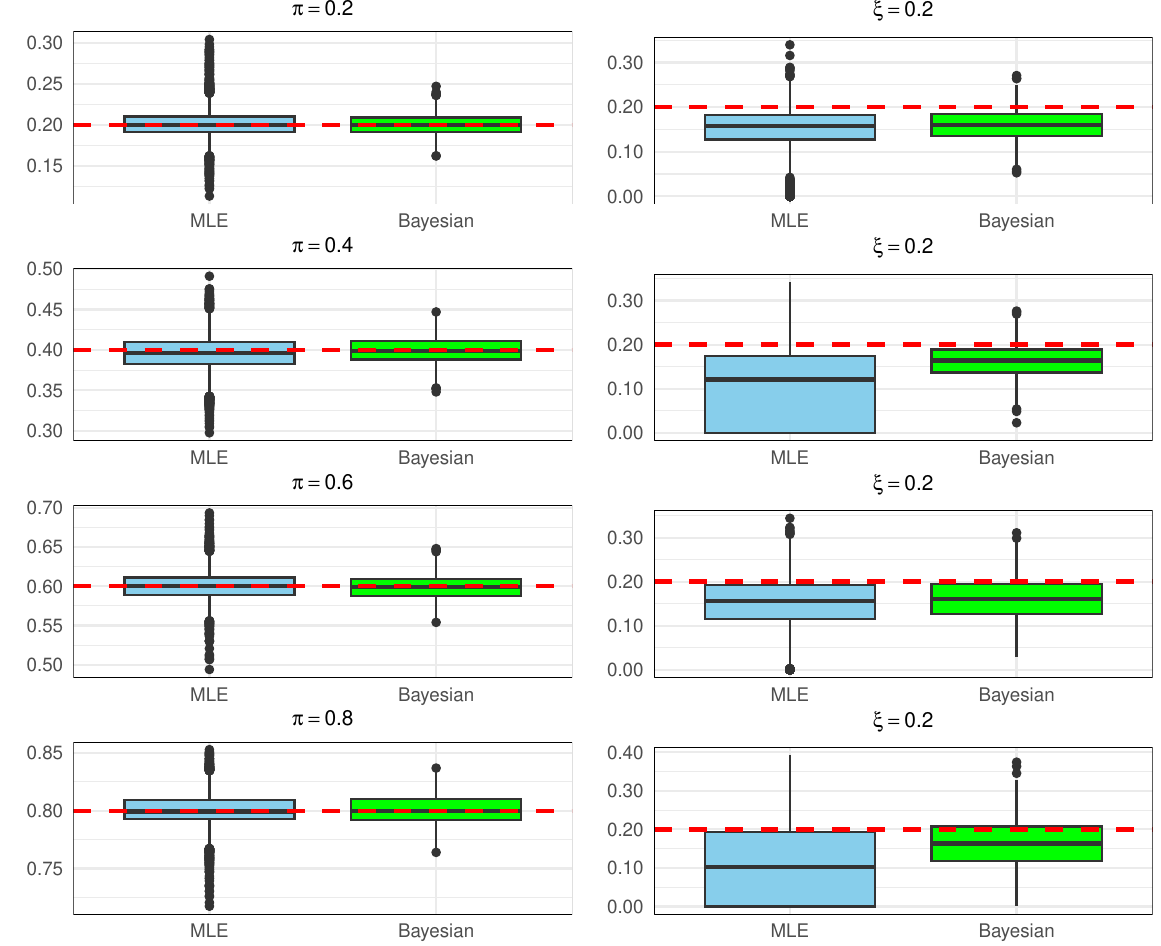}}}\hspace{1pt}
\subfloat[Model $M_3$]{%
\resizebox*{7.5cm}{!}{\includegraphics{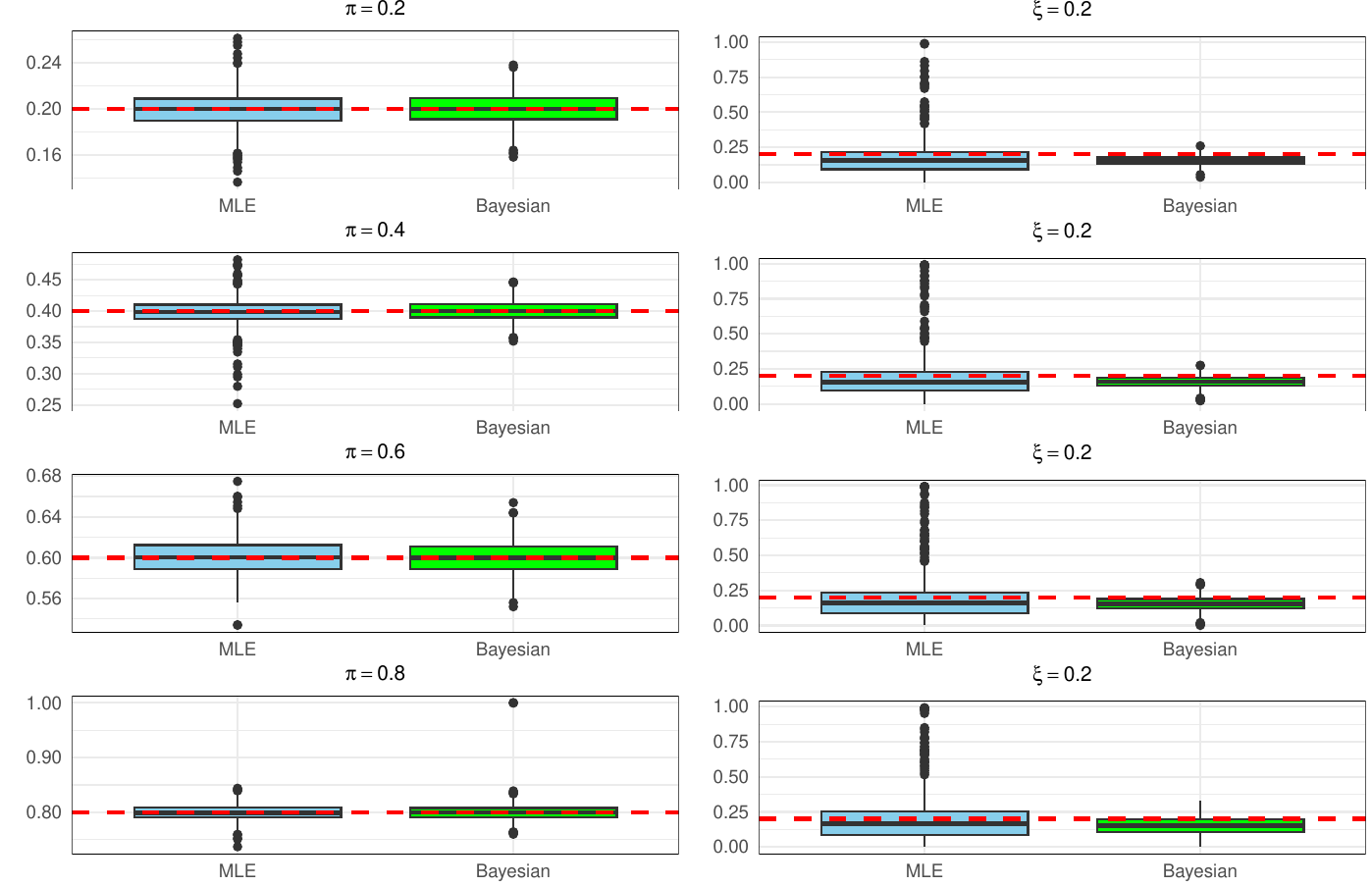}}}
\caption{The box plots illustrate the maximum likelihood estimates and Bayesian estimates of the model parameters for models $M_1$ and $M_3$. These results are based on a sample size of $n = 1000$ and $10^4$ replications, conducted under varying parameter settings with $\pi$ ranging from 0.2 to 0.8 and a fixed $\xi$ value of 0.2, using the ZIGEV distribution.}\label{FIG-ZIGEV}  
\end{figure}
\begin{table}[ht]
    \centering
    \caption{Root mean square errors for Model 1 and Model 3 parameters estimates from ZIGEV through classical and Bayesian when sample size $n=1000$ and $10^4$ are used. }
    \begin{tabular}{cccccc}
        \toprule
        \multicolumn{6}{c}{$\mathcal{W}(u;\Psi)=u^{\kappa}$, $\Psi=\kappa >0$ when $n=1000$} \\ 
        \midrule
        & \multicolumn{2}{c}{RMSE}  & & \multicolumn{2}{c}{RMSE} \\
        \cmidrule(lr){2-3} \cmidrule(lr){5-6} 
        $\pi$ & $\text{MLE}$ & $\text{Bayes}$ &  $\xi$ & $\text{MLE}$ & $\text{Bayes}$\\
        \midrule
        0.2   & 0.016 & 0.013 &  0.2 & 0.082 & 0.053 \\
        0.4  & 0.023 & 0.016 &   0.2 & 0.135 & 0.053 \\
        0.6   & 0.017 & 0.015 &  0.2 & 0.085 & 0.063 \\
        0.8  & 0.014 & 0.013 &   0.2 & 0.138 & 0.079 \\
        \midrule
       \multicolumn{6}{c}{$\mathcal{W}(u;\Psi)=[1-B_{\delta}\{(1-u)^{\delta}\}]^{\kappa/2}$, $\Psi=(\delta, \kappa)$ 
       } \\
        \midrule
        0.2   & 0.013 & 0.013 &  0.2 & 0.119 & 0.056 \\
        0.4  & 0.019 & 0.016 &   0.2 & 0.143 & 0.060 \\
        0.6   & 0.017 & 0.016 &  0.2 & 0.148 & 0.068 \\
        0.8  & 0.013 & 0.016 &   0.2 & 0.165 & 0.084 \\
        \bottomrule
    \end{tabular}\label{RMSE-ZIGEV}
\end{table}

The models $M_1$ and $M_3$ are fitted to the synthetic data, and the results of the simulation experiments are presented in Figure~\ref{FIG-ZIGEV} as boxplots. Model $M_2$ is not fitted to the synthetic data because it lacks flexibility in capturing the bulk of the distribution and tends to impose excessive contraction around the central part of the data \citep{naveau2016modeling}. The parameter $\pi$ is correctly estimated by both methods, no matter what proportion of zero values is considered. On the other hand, the shape parameter $\xi$ in both models is closer to the actual value and has a smaller RMSE (see Table~\ref{RMSE-ZIGEV}) when estimated through the Bayesian method. At the same time, it shows more variation for model $M_1$ when estimated through MLE settings. In model $M_3$, the shape parameter estimates are similar in both methods, but again, a slight deviation is observed in MLE. 

Overall, Figure~\ref{FIG-ZIGEV} illustrates the variability and accuracy of the parameter estimates when data is simulated from other models with different zero-inflation levels. This highlights that the Bayesian method consistently provides more stable shape parameter estimates $\xi$ even when the zero inflation is too high. Therefore, stability is crucial for accurately modeling extreme events, especially when zero inflation is significant.

\section{Real Applications}\label{realapp}

\subsection{Study Area}\label{Study Area}
\begin{figure}[H]
\centering
\subfloat[Map of study sites ]{%
{\includegraphics[width=7cm, height=7cm]{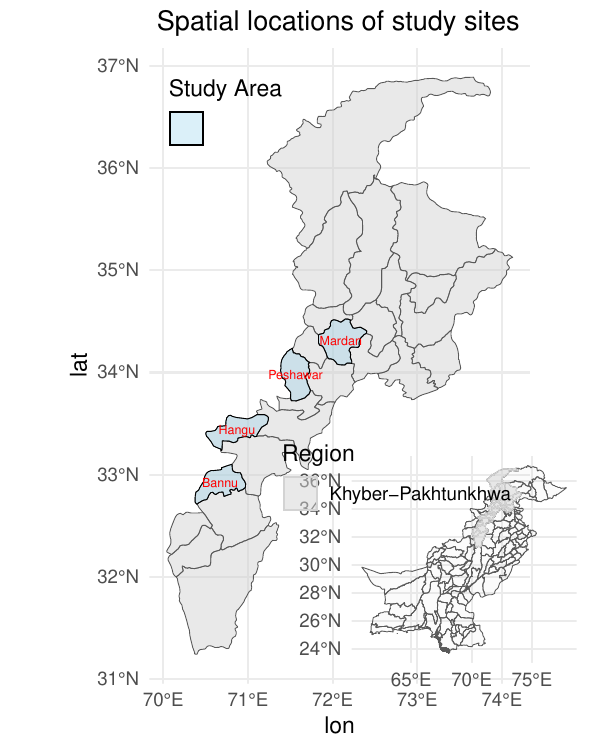}}}\hspace{5pt}
\subfloat[ Dry vs wet events]{%
{\includegraphics[width=7cm, height=7cm]{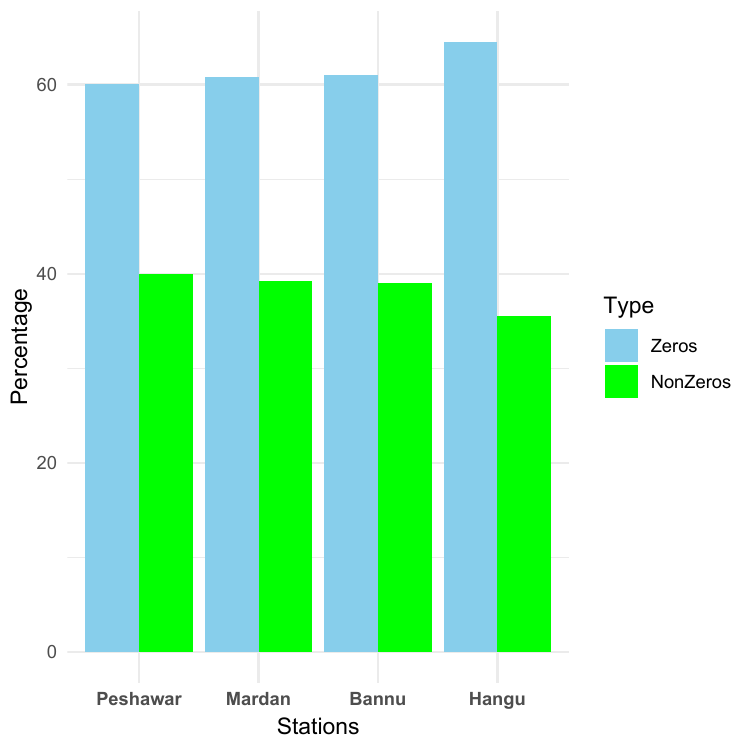}}}
 \caption{\textbf{(a)} Spatial locations of considered study sites, \textbf{(b)} dry vs wet precipitation events. }
	\label{fig:zero-data}
\end{figure}

Khyber Pakhtunkhwa (KP) is Pakistan's third-largest province, covering 101,741 km² area, which is $12.5\%$ of the nation's total land.
KP is situated in the northwest and shares borders with Afghanistan to the north and west, Gilgit-Baltistan to the northeast, Azad Jammu and Kashmir to the east, and the eastern regions of Punjab and Balochistan. The province is located between longitudes 69°E to 74°E and latitudes 31°N to 36°N. The temperature and precipitation patterns vary due to its diverse geography, including plains and river basins in the south and the high Hindu Kush mountains in the north. The northern region has a lot of winter precipitation, particularly in the area near the Hindu Kush. On the other hand, Peshawar and the rest of the southern region experience hot summers and comparatively dry winters due to moderate to low rainfall.
Occasionally, monsoon rains affect northern areas, while summers are generally dry. This variability in precipitation has a significant impact on the province's agriculture, water resources, and climate resilience. We consider the daily average precipitation data from 2015 to 2024 for four different stations: Peshawar, Mardan, Bannu, and Hangu. For the spatial locations of the considered stations, see the left panel of Figure~\ref{fig:zero-data}. Data are downloaded from the Power NASA website \textcolor{red}{https://power.larc.nasa.gov/data-access-viewer/}. Every third value was retained from the data to overcome the temporal dependency. Later, we extracted the data for November, December, January, and February (the winter season), which contained more dry events. \answer{We additionally replaced the values below $0.1 mm$ by zero, considering it a dry day event.}  In our proposal, we are particularly interested in modeling excessive dry events (zero inflation) along with the low, moderate, and extreme wet events (see Figure~\ref{fig:zero-data}, right panel).

\begin{figure}[h]
\centering
\subfloat[Peshawar ]{%
\resizebox*{5.5cm}{!}{\includegraphics{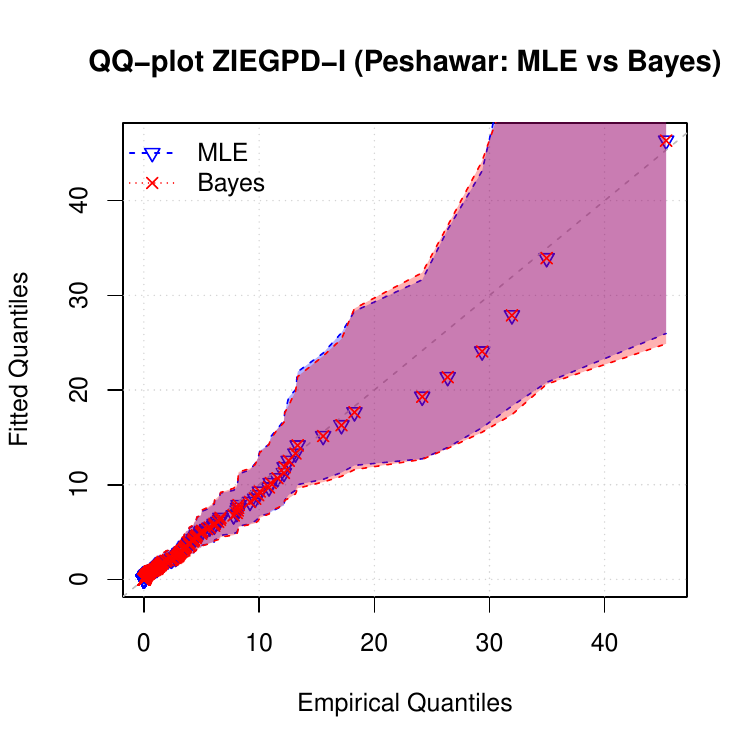}}}
\subfloat[ Mardan]{%
\resizebox*{5.5cm}{!}{\includegraphics{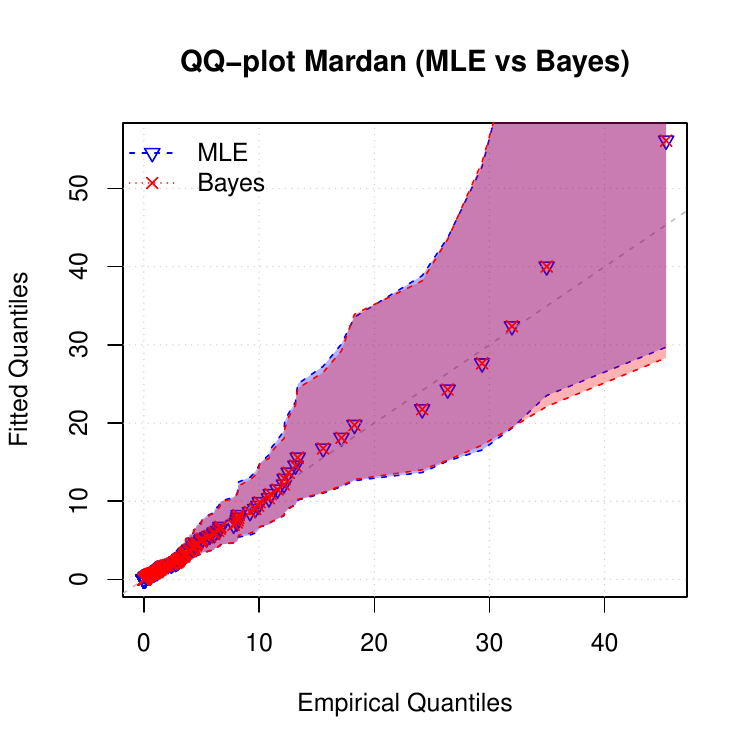}}}
\subfloat[Bannu ]{%
\resizebox*{5.5cm}{!}{\includegraphics{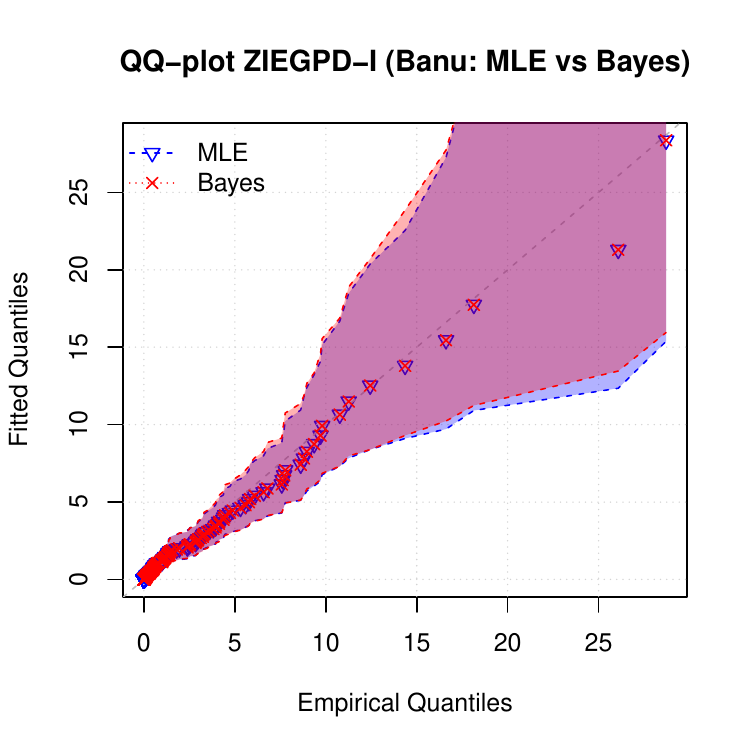}}}\\
\subfloat[Peshawar ]{%
\resizebox*{5.5cm}{!}{\includegraphics{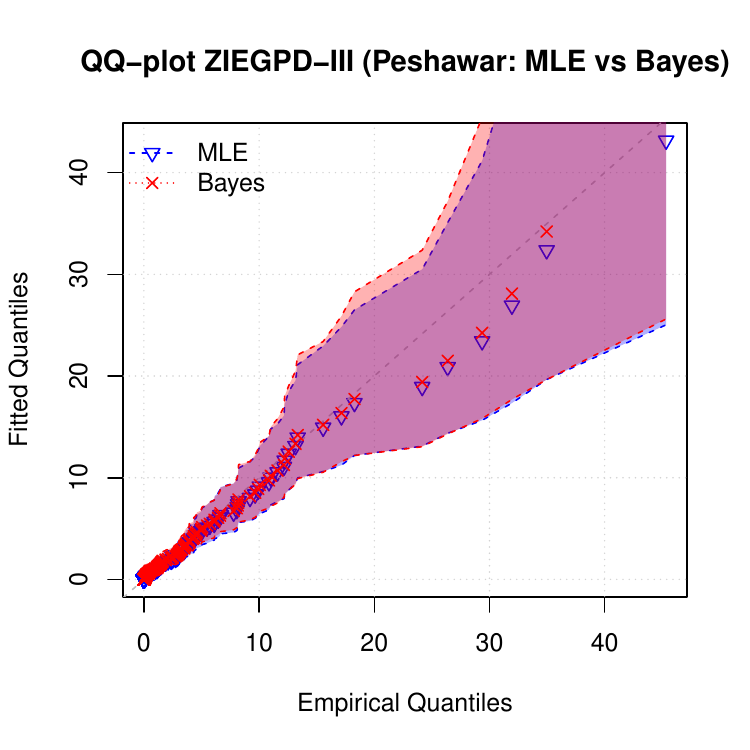}}}
\subfloat[Mardan ]{%
\resizebox*{5.5cm}{!}{\includegraphics{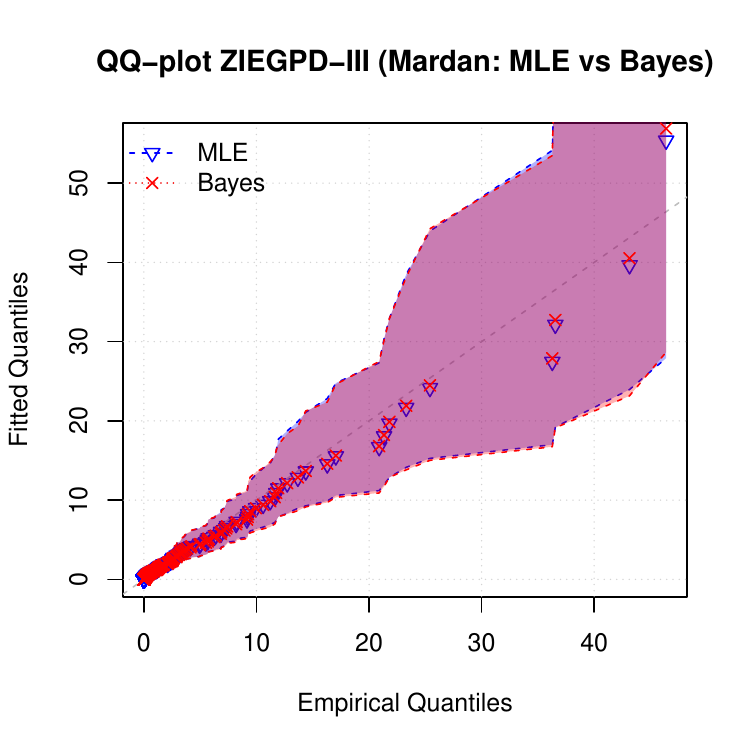}}}
\subfloat[Bannu ]{%
\resizebox*{5.5cm}{!}{\includegraphics{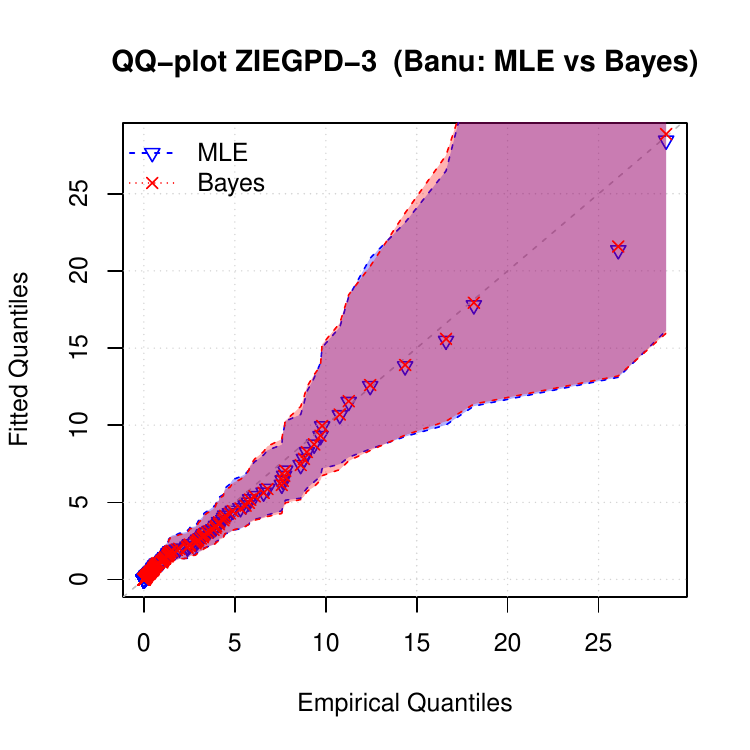}}}
 \caption{QQ-plots of fitted model to stations Peshawar, Mardan and Bannu: Model $M_1$ in (Top panel) and Model $M_3$ in (bottom panel). }
	\label{fig:ZIGEPD-real}
\end{figure}

\begin{figure}[h]
\centering
\subfloat[ZIEGPD-I ]{%
\resizebox*{5cm}{!}{\includegraphics{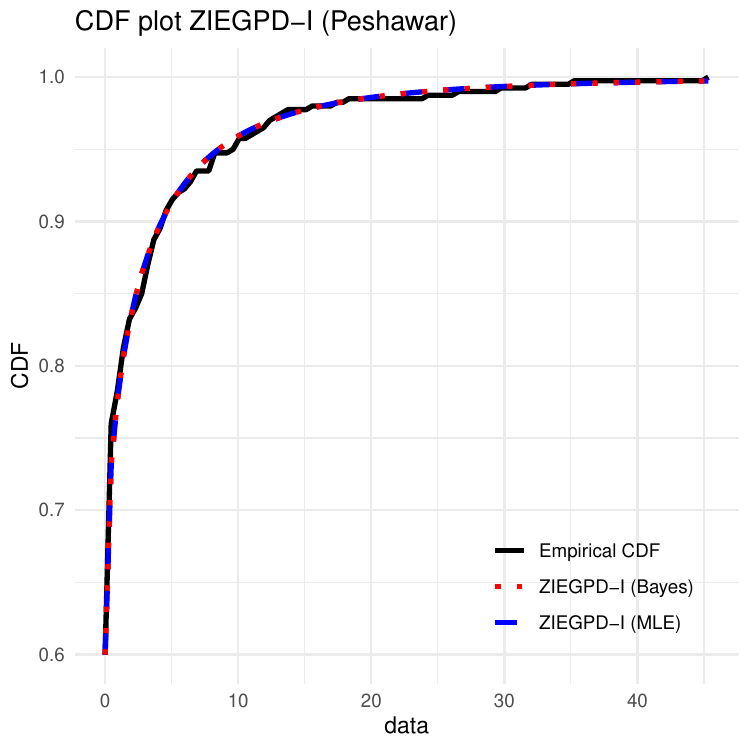}}}
\subfloat[ZIEGPD-I ]{%
\resizebox*{5cm}{!}{\includegraphics{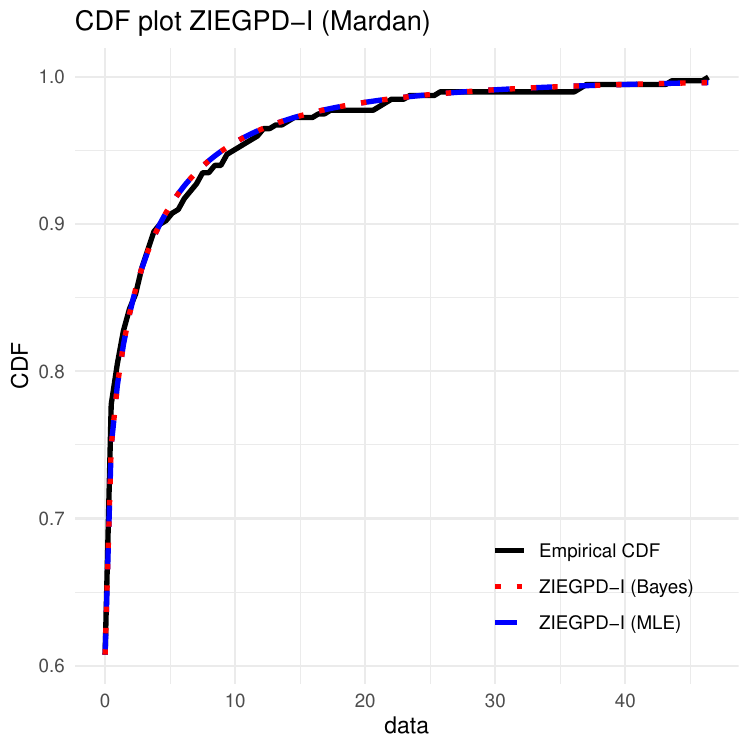}}}
\subfloat[ZIEGPD-I ]{%
\resizebox*{5cm}{!}{\includegraphics{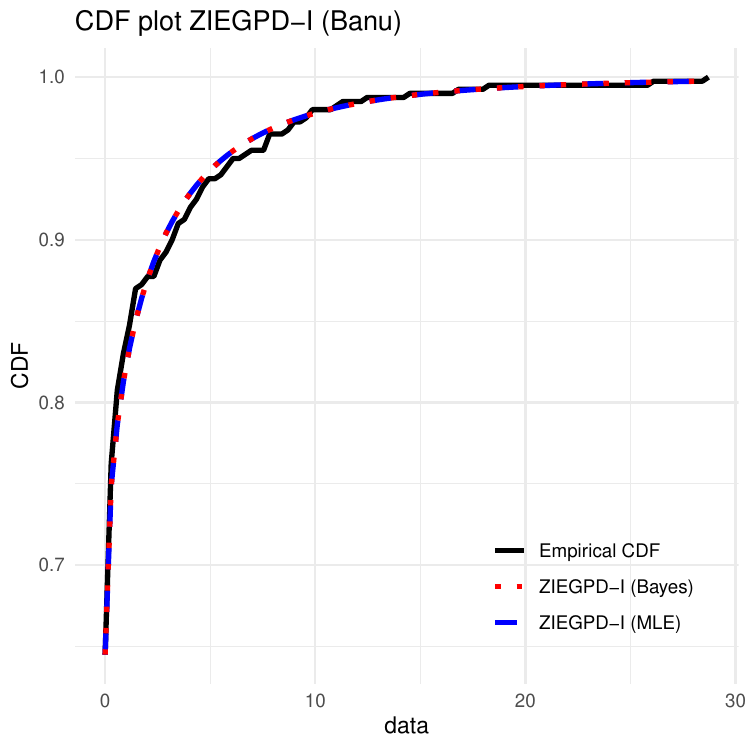}}}\\
\subfloat[ZIEGPD-III ]{%
\resizebox*{5cm}{!}{\includegraphics{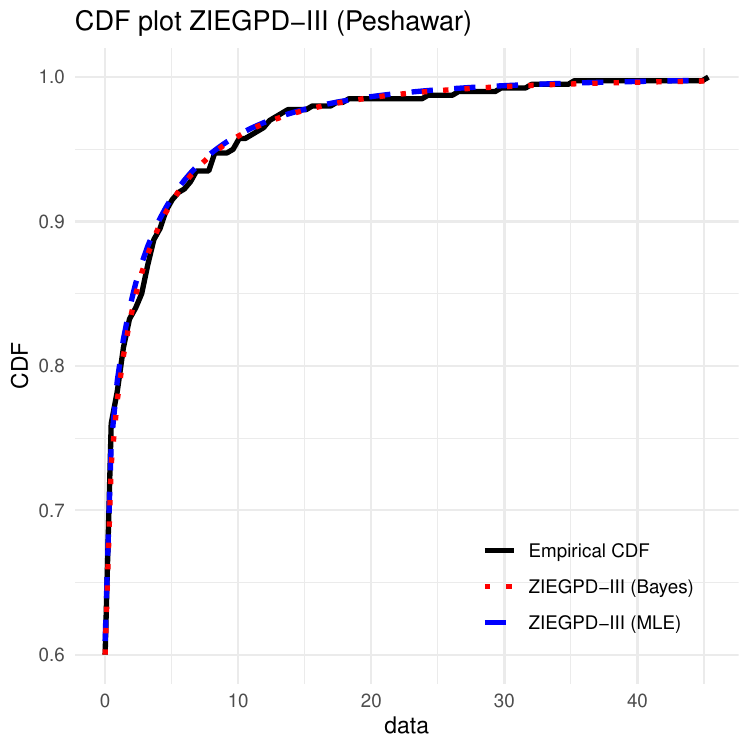}}}
\subfloat[ZIEGPD-III ]{%
\resizebox*{5cm}{!}{\includegraphics{Figures/CDF-m1-Mardan.pdf}}}
\subfloat[ZIEGPD-III ]{%
\resizebox*{5cm}{!}{\includegraphics{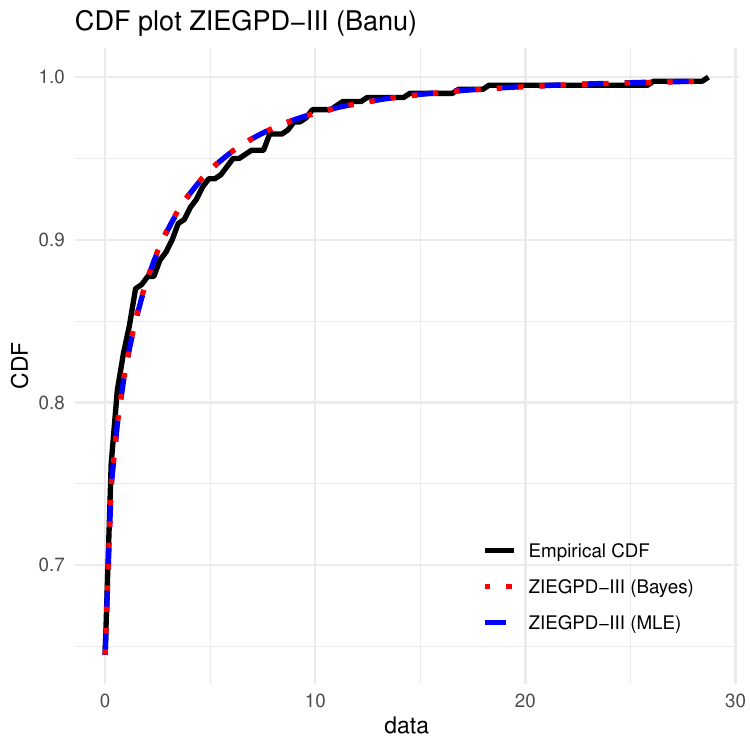}}}
 \caption{CDF-plots of fitted model to stations Peshawar, Mardan and Bannu: Model $M_1$ in (Top panel) and Model $M_3$ in (bottom panel).}
	\label{fig:ZIGEPD-real-cdf}
\end{figure}

The ZIEGPD ($M_1, M_2$ and $M_3$) models were separately fitted to zero-inflated rainfall intensities of each using both ML and Bayesian methods. Model $M_1$ provided the best fit across both approaches. Model $M_3$ also performed well and was comparable to model $M_1$, but its skewness parameter $\delta$ was sometimes poorly estimated in MLE. Model $M_2$ lacked flexibility in the lower tail, consistently making it the weakest performer. 
Therefore, the QQ-plot and CDF-plot of models $M_1$ and $M_3$  for stations (Peshawar, Mardan, and Bannu) are presented in Figure~\ref{fig:ZIGEPD-real} and Figure~\ref{fig:ZIGEPD-real-cdf}, while the results of Hangu station are given in Figure~\ref{fig:ZIGEPD-real-sup1}. As noted, Model 
$ M_2$ performed the worst for all stations; the results of model $M_2$ are also included in SM (see e.g., 
 Figures~\ref{fig:ZIGEPD-real-qq-sup}, \ref{fig:ZIGEPD-real-cdf-sup}). \answer{As observed in both the simulation and real-data analyses, the model $M_2$
 exhibits poor performance, primarily due to its limited flexibility in capturing the lower tail, even in the presence of a substantial proportion of zeros. We nevertheless included this model in our study to serve as a benchmark: despite incorporating a zero-inflated component, its shortcomings remain evident, similar to~\citet{naveau2016modeling}, thereby highlighting the benefits of more flexible alternatives, for instance, models $M_2$ and $M_3$.  }

Table~\ref{Table_real} reports the estimated parameters obtained by the ML method (with 95\%-conﬁdence intervals constructed from 1000 nonparametric bootstrap replicates) and the Bayesian method with 95\% credible intervals. The estimates of parameter $\kappa$ for $M_1$ are almost similar for both methods in all stations; Bayesian estimates for model $M_3$ are slightly smaller than MLE in the first station Peshawar, while for the remaining stations, Bayesian estimates are slightly larger than MLE. The estimated skewness parameter $\delta$ for Peshawar is unstable in both methods. Unlike the Bayesian estimates for the remaining stations, the MLE estimates of \(\delta\) exhibit stability. The reason behind this instability of \(\delta\) is discussed in the simulation study. Individually, the shape parameter \(\xi\) and zero inflation parameter \(\pi\) show similar estimates across both methods in each separate station.

\begin{sidewaystable}
\centering
\scriptsize
\caption{Models $M_1$ and $M_3$ fitted through MLE and Bayesian paradigms to stations (Peshawar, Mardan, Bannu, and Hangu).}
\label{Table_real}
\begin{tabular}{ll lccccc}
\toprule
Station & Method & Model & $\kappa$ & $\delta$ & $\sigma$ & $\xi$ & $\pi$ \\
\midrule

\multirow{6}{*}{Peshawar} 
    & \multirow{3}{*}{MLE} 
        & $M_1$ & 0.4568 (0.3955, 0.5560) & - & 4.9095 (2.2781, 8.0999) & 0.3281 (6.95$\times10^{-6}$, 0.8585) & 0.5999 (0.5475, 0.6475)\\
    &   & $M_3$ & 0.7645 (0.3968, 0.8288) & 92.2779 (0.0005, 190.9993) & 5.8624 (3.3922, 9.3927) & 0.2608 (1.28$\times10^{-6}$, 0.5403) & 0.6094 (0.5522, 0.6486) \\
\cmidrule{2-8}
    & \multirow{3}{*}{Bayesian} 
        & $M_1$ & 0.4570 (0.3645, 0.5762) & - & 4.9067 (2.8377, 7.9964) & 0.3282 (0.0456, 0.7874) & 0.6000 (0.5506, 0.6467) \\
    &   & $M_3$ & 0.6625 (0.5079, 0.9013) & 86.8341 (24.3858, 184.0118) & 7.0006 (3.0222, 11.7988) & 0.3165 (1.18$\times10^{-6}$, 0.6462) & 0.6000 (0.5475, 0.6445) \\

\midrule
\multirow{6}{*}{Mardan} 
    & \multirow{3}{*}{MLE} 
        & $M_1$ & 0.3989 (0.3452, 0.4980) & - & 5.7098 (2.3001, 10.3763) & 0.3703 (2.6$\times10^{-6}$, 0.9899)  & 0.6075 (0.5574, 0.6575) \\
    &   & $M_3$ & 0.3981 (0.3480, 0.7329) & 1.2852 (2.6$\times10^{-5}$, 0.9999) & 6.1497 (2.9834, 11.3623) & 0.35681 (1.4$\times10^{-6}$, 0.8340) & 0.6075 (0.5550, 0.6558) \\
\cmidrule{2-8}
    & \multirow{3}{*}{Bayesian} 
        & $M_1$ & 0.3990 (0.2987, 0.4561) & - & 5.7087 (4.0796, 11.7988) & 0.3703 (0.21$\times10^{-6}$, 0.6788) & 0.6075 (0.5599, 0.6540) \\
    &   & $M_3$ & 0.5648 (0.4579, 0.7141) & 117.3311 (43.6848, 245.4271) & 9.8552 (5.2593, 14.8797) & 0.3494 (1.02$\times10^{-6}$, 0.5547) & 0.6075 (0.5621, 0.6505) \\

\midrule
\multirow{6}{*}{Bannu} 
    & \multirow{3}{*}{MLE} 
        & $M_1$ & 0.4772 (0.4032, 0.6304) & - & 3.6156 (1.3766, 5.9828) & 0.2681 (1.34$\times10^{-6}$, 0.8212) & 0.6450 (0.5964, 0.6898) \\
    &   & $M_3$ & 0.4773 (0.4070, 0.6936) & 0.8852 (4.12$\times10^{-6}$, 20.4997) & 3.5023 (0.9387, 6.5650) & 0.2730 (1.20$\times10^{-6}$, 0.7446) & 0.6450 (0.5998, 0.6923) \\
\cmidrule{2-8}
    & \multirow{3}{*}{Bayesian} 
        & $M_1$ & 0.4772 (0.3668, 0.6168) & - & 3.6129 (1.9140, 6.3968) & 0.2686 (0.12$\times10^{-6}$, 0.8046) & 0.6450 (0.5962, 0.6880) \\
    &   & $M_3$ & 0.6487 (0.4639, 0.8189) & 58.7328 (10.2062, 170.7158) & 6.1288 (4.4593, 8.7845) & 0.2720 (3.28$\times10^{-6}$, 0.8412) & 0.6450 (0.5998, 0.6865) \\

\midrule
\multirow{6}{*}{Hangu} 
    & \multirow{3}{*}{MLE} 
        & $M_1$ & 0.4437 (0.3867, 0.5649) & - & 3.9883 (1.3808, 6.5347) & 0.3819 (0.0071, 0.9899) & 0.6099 (0.5613, 0.6551) \\
    &   & $M_3$ & 0.4426 (0.3899, 0.8688) & 1.1696 (1.08$\times10^{-7}$, 98.8499) & 4.1883 (0.4448, 8.0281) & 0.3737 (2.92$\times10^{-6}$, 0.9998) & 0.6100 (0.5631, 0.6551) \\
\cmidrule{2-8}
    & \multirow{3}{*}{Bayesian} 
        & $M_1$ & 0.4437 (0.3529, 0.5766) & - & 3.9875 (2.0228, 6.7944) & 0.3821 (0.0561, 0.9724) & 0.6100 (0.5628, 0.6580) \\
    &   & $M_3$ & 0.5996 (0.4217, 0.8781) & 96.1587 (26.8966, 194.4160) & 7.9169 (2.8348, 9.7571) & 0.3585 (1.01$\times10^{-6}$, 0.7981) & 0.6100 (0.5643, 0.6528) \\

\bottomrule
\end{tabular}
\end{sidewaystable}

Table~\ref{return-levels} presents the estimated return levels for different return periods (5, 10, 20 years) at four rainfall stations using MLE and Bayesian estimation methods applied to two models, \( M_1 \) and \( M_3 \). The return period, also known as the average recurrence interval (\( T \) years), is defined as \( T = \frac{1}{P} \), where \( P \) is the probability of a rainfall event occurring \answer{in any given year}. It represents the expected time interval between occurrences of an extreme rainfall event of a particular magnitude or greater. The corresponding return level (\( X_T \)) represents the amount of rainfall expected to be exceeded once, on average, in that return period. 

\answer{For example, at the Peshawar station, the MLE method with models \( M_1 \) (resp. $M_3$) estimates a 10-year return level of 4.2433 mm (resp. 4.0101 mm), implying that in any given year there is a 10\% probability (1/10) that rainfall will exceed this amount. Similarly, the 20-year return level is 8.5123 mm for \( M_1 \) and 8.3478 mm for \( M_3 \), which correspond to a 5\% annual exceedance probability (1/20). At shorter return periods (e.g., 5 years), both 
\( M_1 \) and \( M_3 \) 
provide very similar estimates, indicating that the choice of model has a limited influence on lower return levels. However, as the return period increases, differences between the models become more evident, reflecting greater sensitivity of extreme rainfall estimates to model specification at higher quantiles.}

\begin{table}[h]
\centering
\scriptsize
\caption{\answer{Return levels for the stations (Peshawar, Mardan, Bannu, and Hangu).}}
\label{Table:return-levels}

\begin{tabular}{llcccc}
\toprule
Station & Method & Model & 5 (0.800) & 10 (0.900) & 20 (0.950) \\
\midrule
\multirow{4}{*}{Peshawar} 
    & \multirow{2}{*}{MLE} 
        & $M_1$ & 1.2664  & 4.2433  & 8.5123    \\ 
    &   & $M_3$& 1.0612  &  4.0101 &  8.3478 \\
    \cmidrule{2-6}
    & \multirow{2}{*}{Bayesian} 
        & $M_1$ &1.2662 & 4.2423 & 8.5100  \\
   &   & $M_3$ & 1.0250 & 4.2514 & 9.3581  \\
\midrule
\multirow{4}{*}{Mardan} 
    & \multirow{2}{*}{MLE} 
        & $M_1$ & 1.0843 & 4.2036 & 8.9872  \\
   &   & $M_3$ & 1.0865 & 4.2101 & 8.9839  \\
\cmidrule{2-6}
    & \multirow{2}{*}{Bayesian} 
        & $M_1$ & 1.0841 & 4.2028 & 8.9854  \\
    &   & $M_3$ & 0.9227 & 4.7319 & 11.3624  \\ 
\midrule
\multirow{4}{*}{Bannu} 
    & \multirow{3}{*}{MLE} 
        & $M_1$ & 0.7189 & 2.7532 & 5.6242 \\
    &   & $M_3$& 0.7178 & 2.7508 & 5.6232  \\
\cmidrule{2-6}
    & \multirow{2}{*}{Bayesian} 
        & $M_1$ & 0.7185 & 2.7518 & 5.6223  \\
    &   & $M_3$ & 0.7125 & 2.7298 & 5.6050  \\
\midrule
\multirow{4}{*}{Hangu} 
    & \multirow{2}{*}{MLE} 
        & $M_1$ & 0.9170 & 3.3016 & 6.8749  \\
    &   & $M_3$ & 0.9190 & 3.3097 & 6.8833  \\
   
\cmidrule{2-6}
    & \multirow{2}{*}{Bayesian} 
        & $M_1$ & 0.9165 & 3.3005 & 6.8734  \\
    &   & $M_3$ & 0.9084 & 3.2791 & 6.8674  \\

\bottomrule
\end{tabular}\label{return-levels}
\end{table}

\section{Conclusion}\label{concl}
In this work, we proposed the ZIEGPD by merging the extended generalised Pareto distribution with a ZIMM to model the daily rainfall, including dry days (i.e., zero inflation). The proposed model flexibly captures dry days and low to moderate precipitation while modeling extreme precipitation events using a tail index akin to the GPD. A key strength of the ZIEGPD lies in its ability to seamlessly integrate the modeling of dry days, low and moderate rainfall, and extreme precipitation events. Unlike traditional approaches that require manual threshold selection to identify extremes, the ZIEGPD inherently captures tail behavior through a tail index that aligns to GPD. Therefore, it circumvents the threshold selection limitations commonly associated with extreme value analysis.

The proposed model also supports both frequentist and Bayesian estimation approaches. The parameters were estimated using MLE and Bayesian inference, offering flexibility in inference depending on data availability and modeling goals. Moreover, we outlined a modeling framework that accommodates the inclusion of covariates, enabling the extension of the ZIEGPD to more complex and informative settings. However, the current implementation of the ZIEGPD is limited to a univariate framework, where daily rainfall is modeled independently of other relevant variables. 

In future research, we will focus on extending the ZIEGPD to a multivariate case using a copula framework and also consider a spatiotemporal framework that allows for the integration of covariates. By doing so, the model could significantly improve its predictive power and applicability to a wider range of hydrological and climatological studies, particularly in regions with complex weather dynamics or varying climatic influences.

	\section*{Appendix}\label{appendix}
\renewcommand{\theequation}{A.\arabic{equation}}
\setcounter{equation}{0}
\setcounter{section}{0}


\textbf{Appendix A.}\label{appendixA}
By using CDF given in \eqref{CDFGPD} and $\mathcal{W}(u;\Psi)=u^{\kappa}$, $\Psi=\kappa >0$ for \eqref{den-ziegpd}, we found that the density function of ZIEGP distribution as
\begin{eqnarray}
    h(z|\pi, \sigma, \xi, \kappa)&=&\pi \cdot I_0(z) + (1 - \pi) \cdot \frac{d}{dz} \Bigg[\left(1-\left(1+\xi  z/\sigma\right)_{+}^{-1/\xi}\right)^{\kappa} \Bigg]\nonumber\\
     &=& \pi \cdot I_0(z) + (1 - \pi) \cdot \Bigg[\frac{\kappa}{\sigma} \left(1 - \left(1 + \frac{\xi z}{\sigma}\right)_{+}^{-1/\xi}\right)^{\kappa - 1} \left(1 + \frac{\xi z}{\sigma}\right)_{+}^{-1/\xi - 1}\Bigg]
\end{eqnarray}
By using CDF given in \eqref{CDFGPD} and $\mathcal{W}(u;\Psi)=1-B_{\delta}\{(1-u)^{\delta}\}$, $\Psi=\delta>0$ for \eqref{den-ziegpd}, we found that the density function of ZIEGP distribution as
\begin{eqnarray}
    h(z|\pi, \sigma, \xi, \kappa)&=&\pi \cdot I_0(z) + (1 - \pi) \cdot \frac{d}{dz} \Bigg[{\mathcal{H}}(z)+ \frac{1}{\delta}{\left[\Bar{\mathcal{H}}(z)\right]}^{\delta+1}-\frac{1}{\delta}{\left[\Bar{\mathcal{H}}(z)\right]}\Bigg]\nonumber\\
     &=& \pi \cdot I_0(z) + (1 - \pi)\cdot\Bigg[ \frac{1}{\sigma}\left( 1 + \frac{1}{\delta} - \frac{\delta + 1}{\delta} \left[ \left(1 + \frac{\xi z}{\sigma}\right)_{+}^{-1/\xi}\right]^{\delta} \right)\left(1 + \frac{\xi z}{\sigma}\right)_{+}^{-1/\xi - 1}
     \Bigg]\nonumber\\
\end{eqnarray}
By using CDF given in \eqref{CDFGPD} and $\mathcal{W}(u;\Psi)=[1-B_{\delta}\{(1-u)^{\delta}\}]^{\kappa/2}$, $\Psi=(\delta, \kappa)$ with $\delta>0$  and $\kappa>0$. For \eqref{den-ziegpd}, we found that the density function of the ZIEGP distribution as
\begin{eqnarray}
    h(z|\pi, \sigma, \xi, \kappa)&=&\pi \cdot I_0(z) + (1 - \pi) \cdot \frac{d}{dz} \Bigg[ \left(\mathcal{H}(z) + \frac{1}{\delta}{\left[\Bar{\mathcal{H}}(z)\right]}^{\delta+1}-\frac{1}{\delta}{\left[\Bar{\mathcal{H}}(z)\right]}\right)^{\frac{\kappa}{2}}\Bigg]\nonumber\\  
     &=& \pi \cdot I_0(z) + (1 - \pi)\cdot\Bigg[ \frac{\kappa}{2\sigma}\Bigg[1 - \left(1 + \frac{\xi z}{\sigma}\right)_{+}^{-1/\xi} + \frac{1}{\delta}{\left( \left(1 + \frac{\xi z}{\sigma}\right)_{+}^{-1/\xi}\right)}^{\delta+1} \nonumber\\
&&-\frac{1}{\delta}{ \left(1 + \frac{\xi z}{\sigma}\right)_{+}^{-1/\xi}}\Bigg]^{\frac{\kappa}{2} -1}
     \left( 1 + \frac{1}{\delta} - \frac{\delta + 1}{\delta} \left[ \left(1 + \frac{\xi z}{\sigma}\right)_{+}^{-1/\xi}\right]^{\delta} \right)\left(1 + \frac{\xi z}{\sigma}\right)_{+}^{-1/\xi - 1}
     \Bigg]
\nonumber\\    
\end{eqnarray}
\\
\answer{\textbf{Appendix B.}\label{appendixB}
In case of non-stationary data, we also developed the non-stationary version of the proposed models. It is a semi-parametric regression model that assumes a parametric distribution for the response variable, with the option for the distribution parameters to change according to explanatory variables or random effects. In non-stationary framework, the observation $z_t, t=1, \cdots, n$ and fitted to distribution $ h(z_t|\theta^t)$ with $\theta^t=(\theta_{1t}, \cdots, \theta_{pt} )$ where $p$ is the number of distribution parameters at time $t$.  To relate the parameters of the distribution to time $t$ or vector of covariates $(\boldsymbol{x_i})$ of dimension $p$, we consider a generic structure of additive predictors of the form 
\begin{equation}
\eta_j^{\theta^t}=\beta_{j0}^{{\theta^t}}+ f_{j1}^{{\theta^t}}(\boldsymbol{x}_{ij1})+\dots+ f_{jk}^{{\theta^t}}(\boldsymbol{x}_{ijk})+\dots+f_{jK_j}^{{\theta^t}}(\boldsymbol{x}_{ijK_j}),
\end{equation}
where $\eta_j(\cdot)$ serves as transformers or link functions and rigorously ensures the range of parameters is met as required for each transformation, 
$\beta_{j0}\in \mathbb{R}$ behaves as a global intercept aligned with the term
$j$,
 \( \boldsymbol{x}_{ijk} \) represents the \( k \)-th component within the covariate vector associated with term \( j \) for observation \( i \). The functions \( f_{jk}^{\theta^t}(\cdot) \), parameterized by \( K_j \), delineate the specific covariate effects on the response through term \( j \), with each \( f_{jk}^{\theta^t}(\cdot) \) encompassing flexible formulations either in fixed parametric expressions (e.g., linear or quadratic) or adaptable, non-parametric structures. Each of these functions \( f_{jk}^{\theta^t}(x_{ijk}) \) is approximated by a linear combination of \( J_{kj} \) basis functions \( b_{kjl}(x_{ikj}) \) and regression coefficients \( \beta_{kjl} \in \mathbb{R} \), such that,
\[
f_{jk}(x_{ijk}) \approx \sum_{l=1}^{J_{kj}} \beta_{kjl} b_{kjl}(\boldsymbol{x}_{ikj}).
\]
In matrix notation we can write $ \mathbf{f}_{jk} = \left( f_j(x_{ij1}), \dots, f_j(x_{ijn}) \right)^{'} = \mathbf{Z}_j \boldsymbol{\beta}_j $, where $\mathbf{Z}_j\left[ij, J_{kj}\right]= b_{kjl}(\boldsymbol{x}_{ikj})$ is a design matrix  and  $\boldsymbol{\beta}_j$ are vector of coefficient to be estimated. This approach,
commonly known as distributional regression, allows for a vast variety of covariate effects through regression splines.
}

\section*{Acknowledgement(s)}

Touqeer Ahmad acknowledges support from the Project JUNON.

\section*{Disclosure statement}

The authors declare no conflict of interest.

\section*{Funding}

This study is not receiving any specific funding.



\section*{Data and code availability}
The data and codes prepared for this paper are available for reproducing the results and can be obtained from the corresponding author upon reasonable request.

\bibliographystyle{apalike}
\bibliography{bibliography}


\newpage

\newpage

\renewcommand{\thetable}{S.\arabic{table}}
\setcounter{table}{0}
\renewcommand{\thefigure}{S.\arabic{figure}}
\setcounter{figure}{0}

\renewcommand{\theequation}{S.\arabic{equation}}
\setcounter{equation}{0}
\renewcommand{\thesection}{S.\arabic{section}}
\setcounter{section}{0}
\setcounter{page}{1}
 \begin{center}
 	\section*{Supplementary material for \\ ``Modeling zero-inflated precipitation extremes''}
 \end{center}

\section{Simulation Study}

\begin{figure}[h]
\centering
\subfloat[$M_1$ with $n=500$]{%
\resizebox*{7cm}{!}{\includegraphics{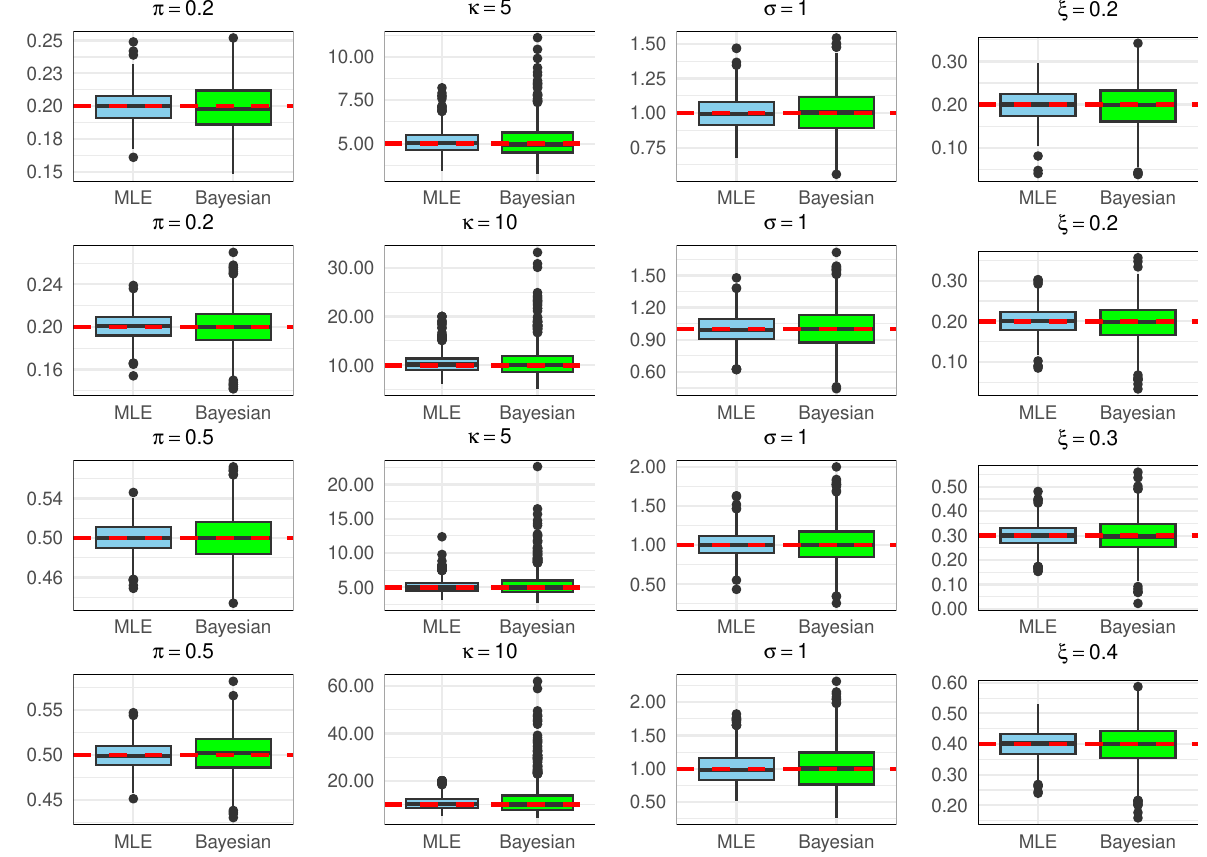}}}\hspace{1pt}
\subfloat[$M_2$ with $n=500$]{%
\resizebox*{7cm}{!}{\includegraphics{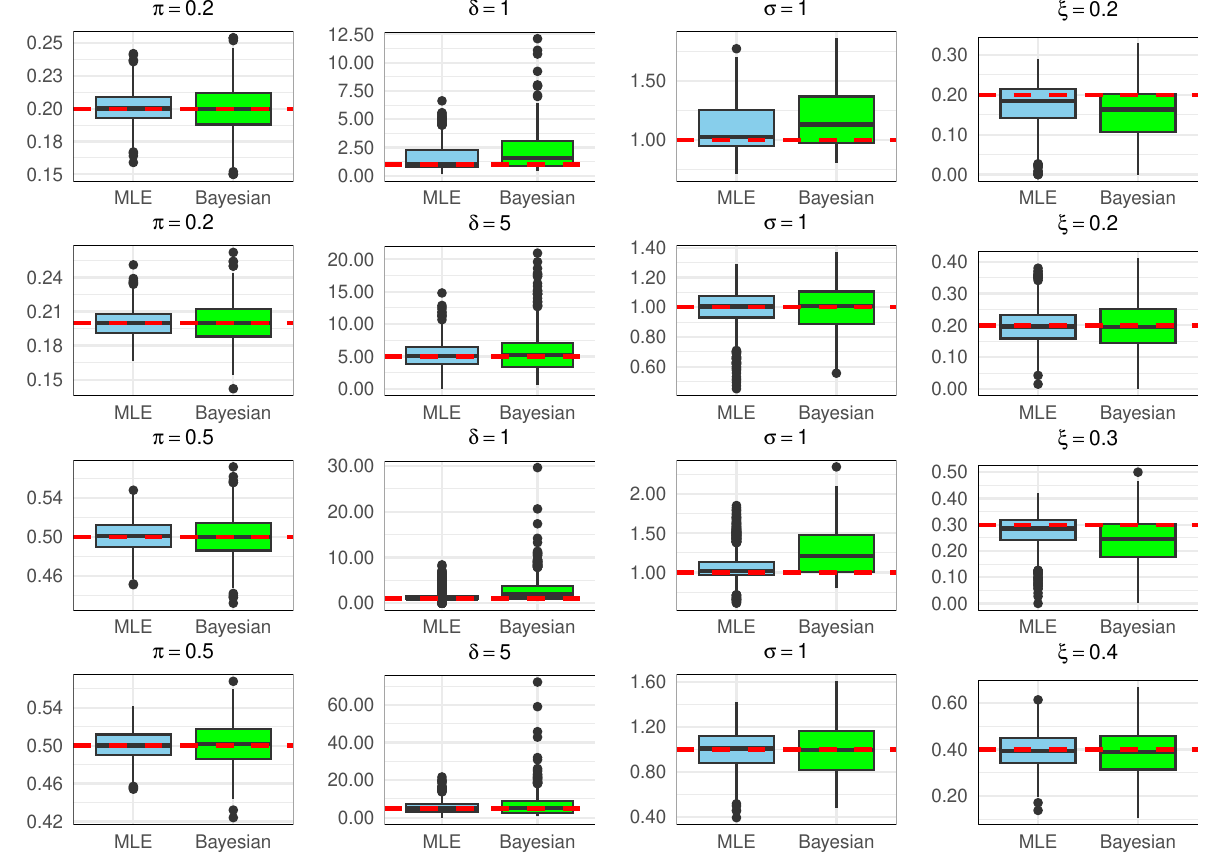}}}\hspace{1pt}
\subfloat[$M_3$ with $n=500$]{%
\resizebox*{7cm}{!}{\includegraphics{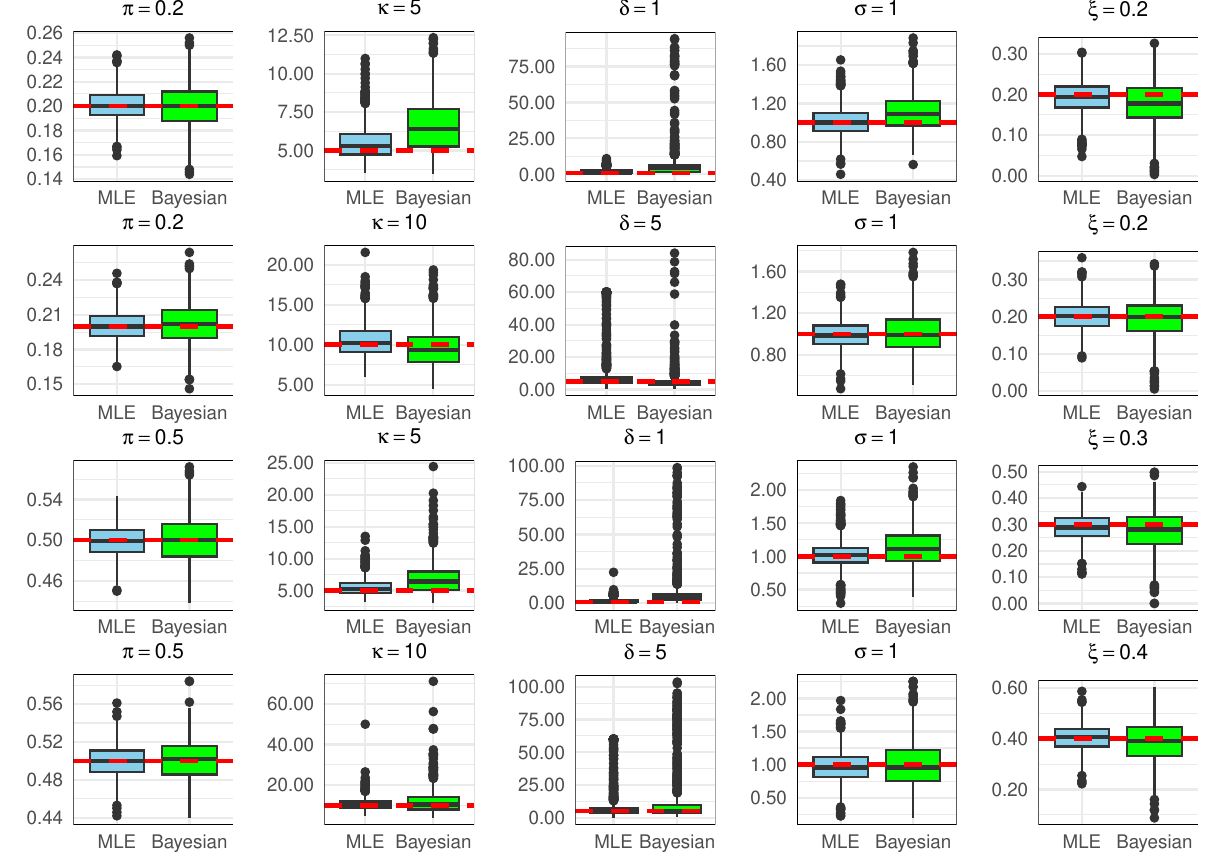}}}\hspace{1pt}
\caption{\answer{Box plots showing the maximum likelihood estimates and Bayesian estimates of model parameters for each model type, based on sample sizes of $n = 500$ with $10^4$ replications under varying parameter settings.}}      
\label{fig:model-based-sim-Sup}
\end{figure}

\begin{table}[]
    \centering
    \caption{Root mean square errors for the parameters of Model 2 estimated through classical and Bayesian when sample size $n=500, 1000$ used. }
    \begin{tabular}{ccccccccccccccccccc}
        \toprule
        \multicolumn{12}{c}{$\mathcal{W}(u;\Psi)=1-B_{\delta}\{(1-u)^{\delta}\}$, $\Psi=\delta>0$ when $n=500$} \\ 
        \midrule
        & \multicolumn{2}{c}{RMSE} & & \multicolumn{2}{c}{RMSE} & & \multicolumn{2}{c}{RMSE} & & \multicolumn{2}{c}{RMSE} \\
        \cmidrule(lr){2-3} \cmidrule(lr){5-6} \cmidrule(lr){8-9} \cmidrule(lr){11-12}
        $\pi$ & $\text{MLE}$ & $\text{Bayes}$ & $\delta$ & $\text{MLE}$ & $\text{Bayes}$ & $\sigma$ & $\text{MLE}$ & $\text{Bayes}$ & $\xi$ & $\text{MLE}$ & $\text{Bayes}$\\
        \midrule
         0.2   & 0.013 & 0.018 & 1 & 1.23 & 1.95 & 1 & 0.22 & 0.29 & 0.2 & 0.06 & 0.08 \\
        0.2   & 0.012 & 0.018 & 5 & 2.05 & 3.10 & 1 & 0.12 & 0.16 & 0.2 & 0.06 & 0.08 \\
        0.5   & 0.016 & 0.022 & 1 & 1.35 & 3.03 & 1 & 0.23 & 0.39 & 0.3 & 0.07 & 0.11 \\
        0.5   & 0.016 & 0.022 & 5 & 3.33 & 6.47 & 1 & 0.19 & 0.23 & 0.4 & 0.08 & 0.11 \\
        \midrule
        \multicolumn{12}{c}{$\mathcal{W}(u;\Psi)=1-B_{\delta}\{(1-u)^{\delta}\}$, $\Psi=\delta>0$ when $n=1000$} \\
        \midrule
        0.2   & 0.013 & 0.012 & 1 & 1.32 & 1.42 & 1 & 0.22 & 0.24 & 0.2 & 0.06 & 0.06 \\
        0.2   & 0.013 & 0.012 & 5 & 2.16 & 2.23 & 1 & 0.13 & 0.12 & 0.2 & 0.06 & 0.06 \\
        0.5   & 0.016 & 0.016 & 1 & 2.12 & 2.28 & 1 & 0.33 & 0.34 & 0.3 & 0.08 & 0.08 \\
        0.5   & 0.016 & 0.016 & 5 & 3.36 & 4.21 & 1 & 0.19 & 0.18 & 0.4 & 0.08 & 0.07 \\
        \bottomrule
    \end{tabular}\label{RMSE-M2}
\end{table}

\begin{table}[ht]
    \centering
    \caption{Root mean square errors for the parameters of Model 3 estimated through classical and Bayesian when sample size $n=500, 1000$ used. }
    \begin{tabular}{ccccccccccccccccccc}
        \toprule
        \multicolumn{15}{c}{$\mathcal{W}(u;\Psi)=[1-B_{\delta}\{(1-u)^{\delta}\}]^{\kappa/2}$, $\Psi=(\delta, \kappa)$ with $\delta>0$  and $\kappa>0$ when $n=500$} \\ 
        \midrule
        & \multicolumn{2}{c}{RMSE} & & \multicolumn{2}{c}{RMSE} & & \multicolumn{2}{c}{RMSE} & & \multicolumn{2}{c}{RMSE} && \multicolumn{2}{c}{RMSE} \\
        \cmidrule(lr){2-3} \cmidrule(lr){5-6} \cmidrule(lr){8-9} \cmidrule(lr){11-12} \cmidrule(lr){14-15}
        $\pi$ & $\text{MLE}$ & $\text{Bayes}$ & $\kappa$ & $\text{MLE}$ & $\text{Bayes}$ & $\delta$ & $\text{MLE}$ & $\text{Bayes}$ & $\sigma$ & $\text{MLE}$ & $\text{Bayes}$ & $\xi$ & $\text{MLE}$ & $\text{Bayes}$  \\
        \midrule
         0.2 & 0.013  & 0.018 & 5 &  1.24 & 2.31 & 1 & 1.90 & 14.21 & 1 & 0.15& 0.22 & 0.2 & 0.04& 0.06 \\
        0.2 & 0.013 & 0.018 & 10 &  2.15 & 2.39 & 5 & 17.43 & 6.52 & 1 & 0.14& 0.19 & 0.2 & 0.04& 0.05 \\
        0.5 & 0.016  &0.022  & 5 &  1.43 & 3.06 & 1 & 2.16 & 16.89 & 1 & 0.21& 0.33 & 0.3 & 0.05& 0.08 \\
        0.5 & 0.017 &0.022  & 10 &  3.07 & 6.35 & 5 & 16.00 & 23.21 & 1 & 0.24& 0.34 & 0.4 & 0.05& 0.08 \\
        \midrule
        \multicolumn{15}{c}{$\mathcal{W}(u;\Psi)=[1-B_{\delta}\{(1-u)^{\delta}\}]^{\kappa/2}$, $\Psi=(\delta, \kappa)$ with $\delta>0$  and $\kappa>0$ when $n=1000$} \\
        \midrule
        0.2 & 0.013  & 0.013 & 5 &  1.16 & 1.94 & 1 & 1.94 & 11.44 & 1 & 0.15& 0.17 & 0.2 & 0.04& 0.05 \\
        0.2 & 0.013 & 0.013 & 10 &  2.14 & 1.86 & 5 & 18.05 & 4.27 & 1 & 0.13& 0.13 & 0.2 & 0.04& 0.04 \\
        0.5 & 0.016  &0.016  & 5 &  1.51 & 2.39 & 1 & 2.03 & 12.62 & 1 & 0.20& 0.25 & 0.3 & 0.05& 0.06 \\
        0.5 & 0.016 &0.016  & 10 &  3.04 & 3.97 & 5 & 15.87 & 22.07 & 1 & 0.24& 0.24 & 0.4 & 0.05& 0.06 \\
        \bottomrule
    \end{tabular}\label{RMSE-M3}
\end{table}

\begin{figure}[h]
\centering
\subfloat[ZIEGPD-I ]{%
\resizebox*{6cm}{!}{\includegraphics{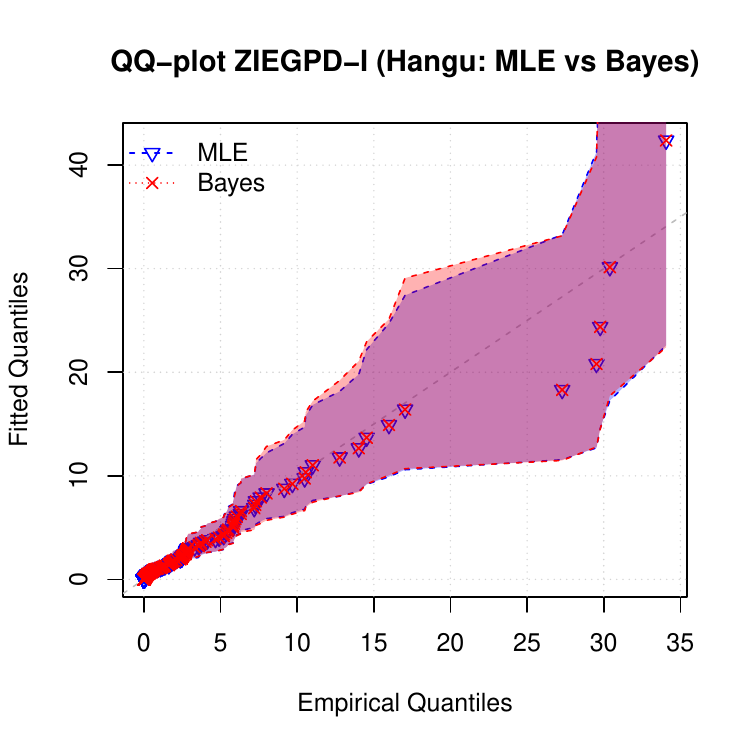}}}\hspace{5pt}
\subfloat[ZIEGPD-III ]{%
\resizebox*{6cm}{!}{\includegraphics{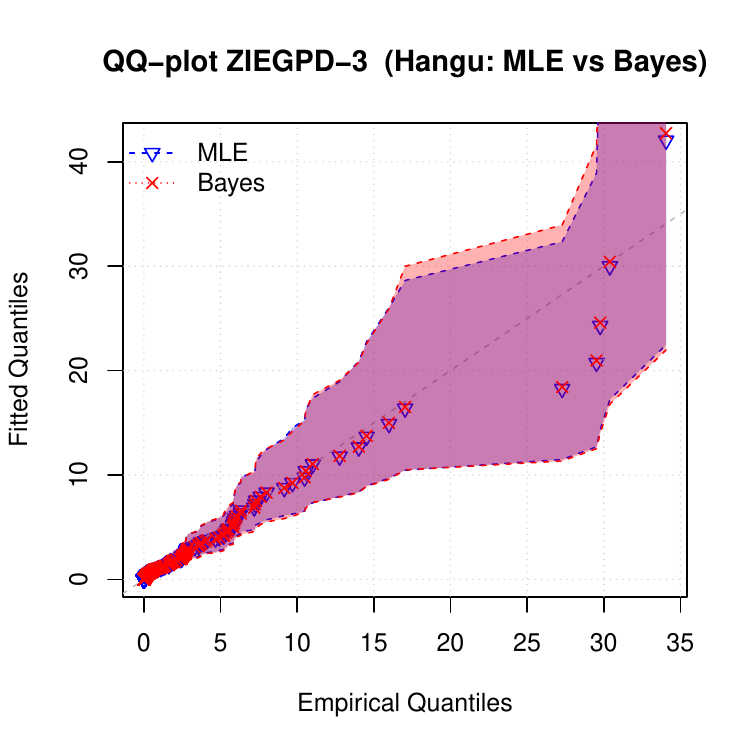}}}\\
\subfloat[ZIEGPD-I ]{%
\resizebox*{6cm}{!}{\includegraphics{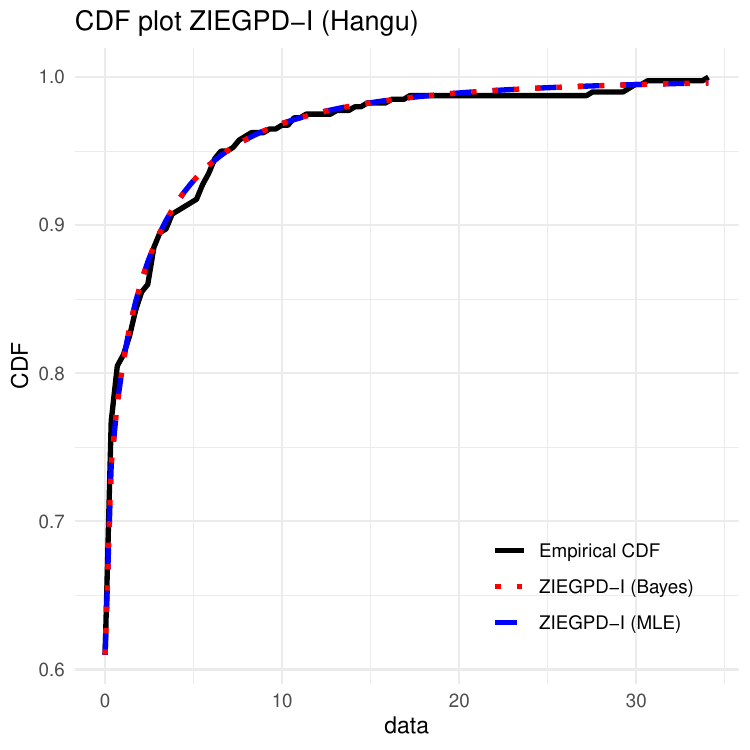}}}\hspace{5pt}
\subfloat[ZIEGPD-III ]{%
\resizebox*{6cm}{!}{\includegraphics{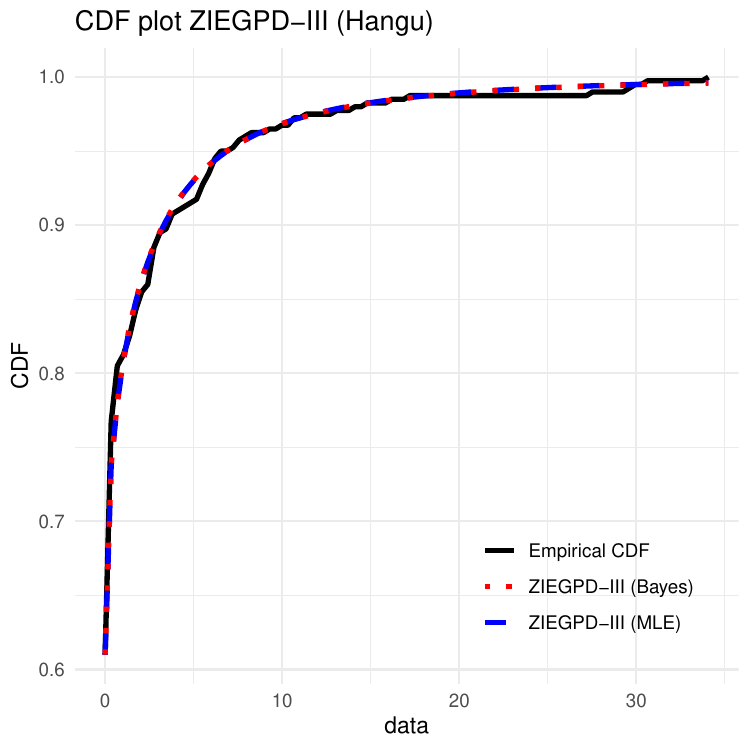}}}
 \caption{QQ-plots of fitted models ($M_1$ and $M_3$) to stations Hangu (top panel), CDF plots (bottom panel).}
	\label{fig:ZIGEPD-real-sup1}
\end{figure}

\begin{figure}[h]
\centering
\subfloat[ZIEGPD-II ]{%
\resizebox*{5cm}{!}{\includegraphics{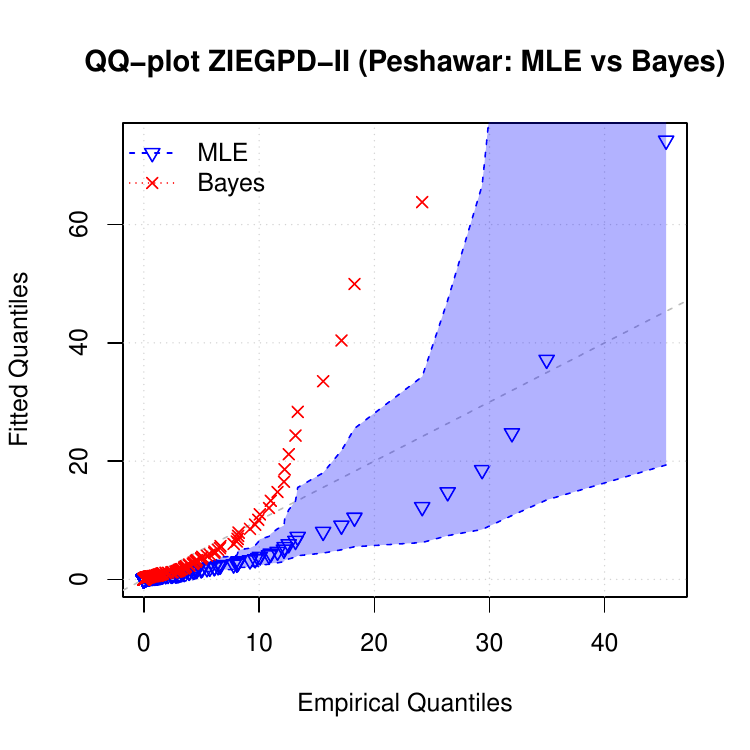}}}
\subfloat[ZIEGPD-II ]{%
\resizebox*{5cm}{!}{\includegraphics{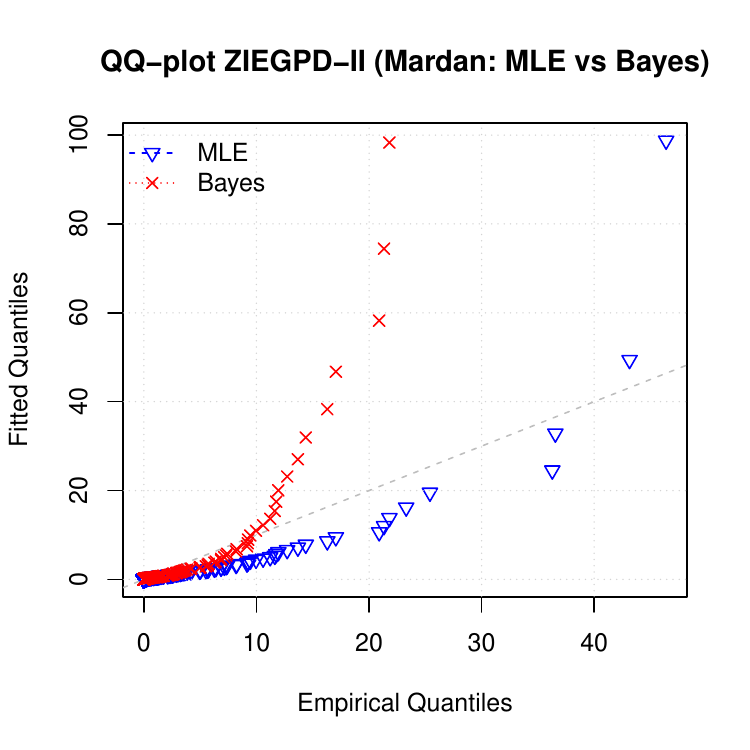}}}\\
\subfloat[ZIEGPD-II ]{%
\resizebox*{5cm}{!}{\includegraphics{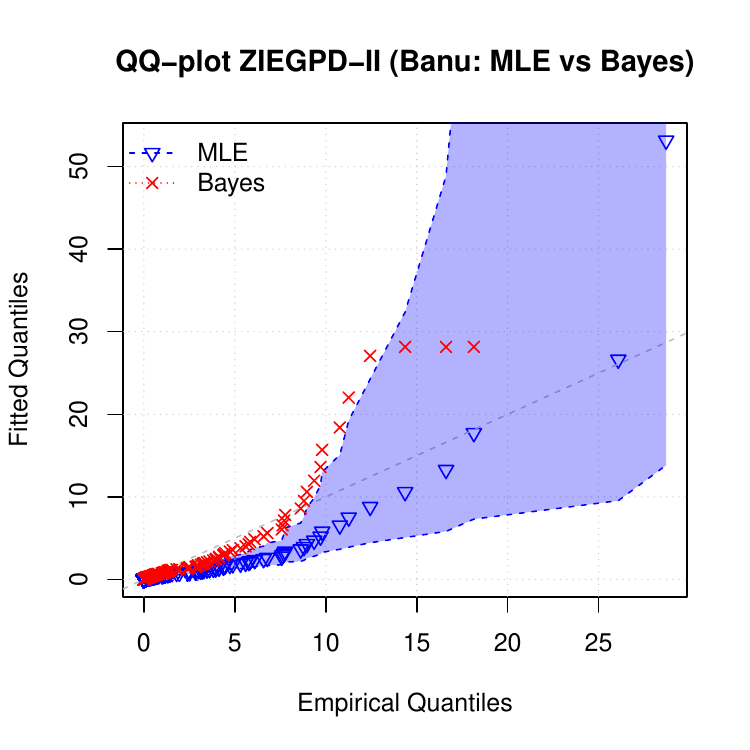}}}
\subfloat[ZIEGPD-II ]{%
\resizebox*{5cm}{!}{\includegraphics{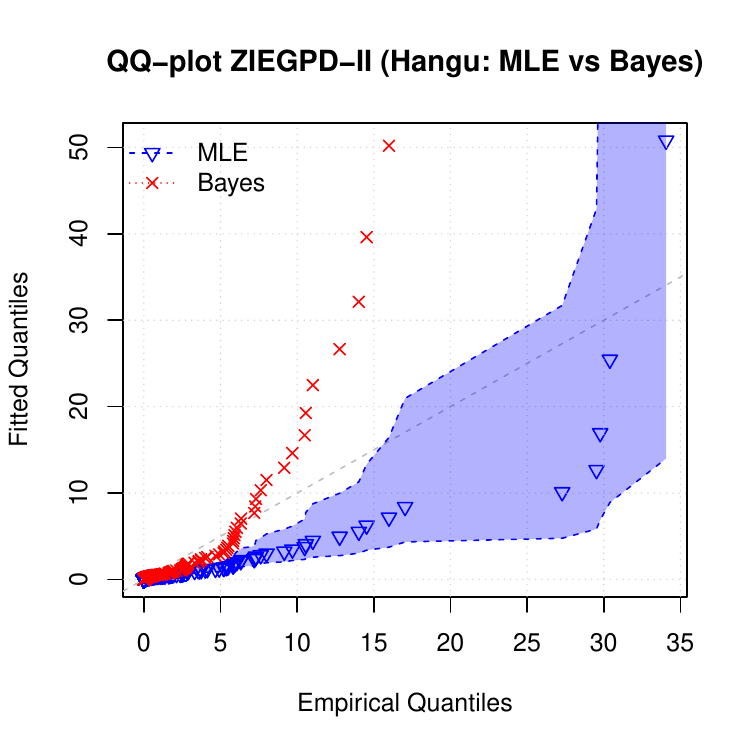}}}
 \caption{QQ-plots plots of fitted model $M_2$ to staions Peshawar, Mardan Bannu and Hangu.}
	\label{fig:ZIGEPD-real-qq-sup}
\end{figure}

\begin{figure}[h]
\centering
\subfloat[ZIEGPD-II ]{%
\resizebox*{5cm}{!}{\includegraphics{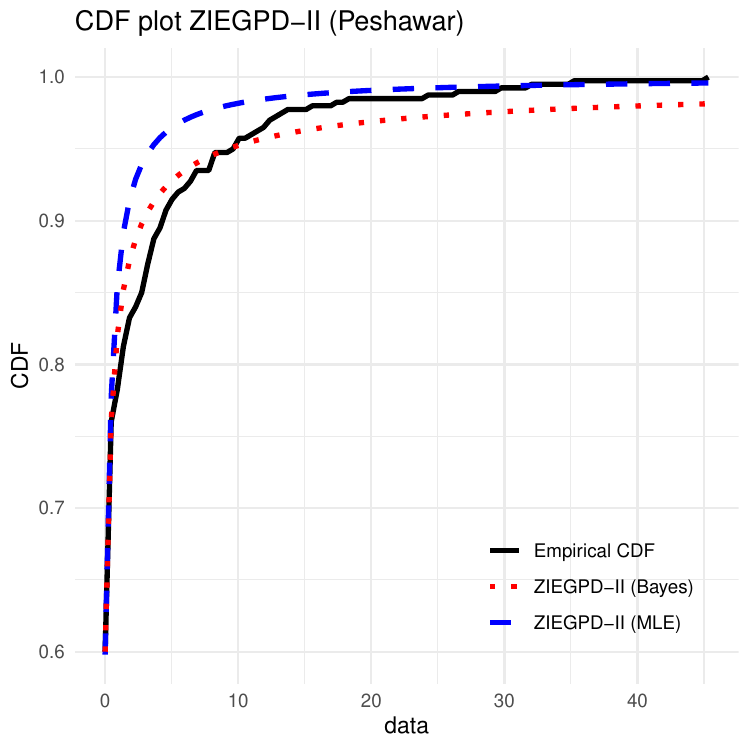}}}
\subfloat[ZIEGPD-II ]{%
\resizebox*{5cm}{!}{\includegraphics{Figures/CDF-m2-mardan.pdf}}}\\
\subfloat[ZIEGPD-II ]{%
\resizebox*{5cm}{!}{\includegraphics{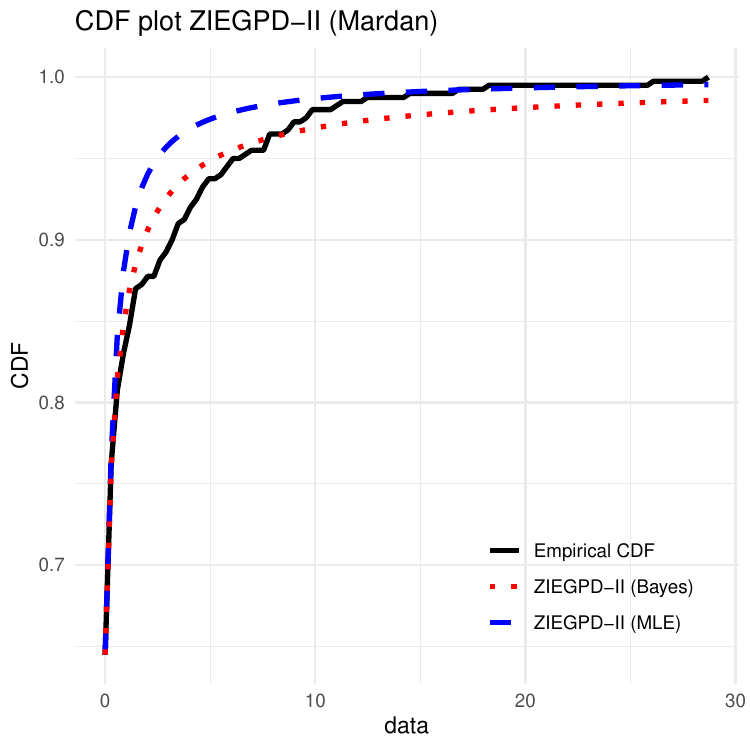}}}
\subfloat[ZIEGPD-II ]{%
\resizebox*{5cm}{!}{\includegraphics{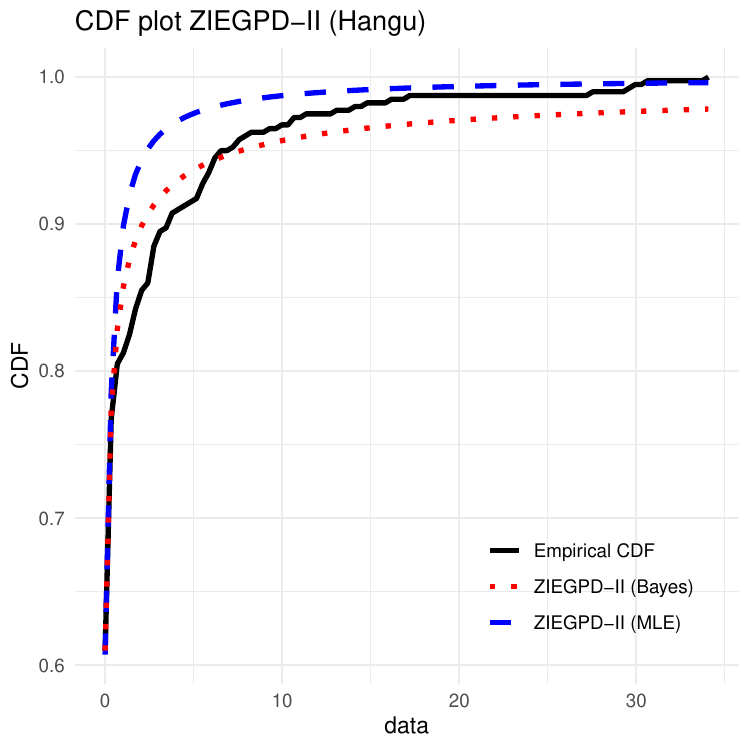}}}
 \caption{CDF plots of fitted model $M_2$ to staions Peshawar, Mardan Bannu and Hangu.}
	\label{fig:ZIGEPD-real-cdf-sup}
\end{figure}


\begin{algorithm}
\caption{Nonparametric Bootstrap for ZIEGPD Confidence Intervals}
\label{alg:Bootstrap_ZIEGPD}
\begin{algorithmic}[1]
\Statex \textbf{Input:} Observed data $\{z_1, \ldots, z_n\}$, number of bootstrap replicates $B$, confidence level $(1-\alpha)$.
\State Fit the ZIEGPD model to $\{z_1, \ldots, z_n\}$ and obtain parameter estimates 
\[
\hat{\theta} = (\hat{\pi}, \hat{\kappa}, \hat{\sigma}, \hat{\xi}),
\]
for example, under Model $M_1$.
\State Initialize bootstrap storage 
\[
\{\pi^{*(b)}, \kappa^{*(b)}, \sigma^{*(b)}, \xi^{*(b)}\}_{b=1}^B.
\]
\For{$b = 1, \ldots, B$}
    \State Draw a bootstrap resample $\{z_1^*, \ldots, z_n^*\}$ from $\{z_1, \ldots, z_n\}$ with replacement.
    \State Fit the ZIEGPD model to $\{z_1^*, \ldots, z_n^*\}$ and obtain
    \[
    \hat{\theta}^{*(b)} = \big(\pi^{*(b)}, \kappa^{*(b)}, \sigma^{*(b)}, \xi^{*(b)}\big).
    \]
    \State Store $\hat{\theta}^{*(b)}$.
\EndFor
\State Construct percentile confidence intervals for each parameter $\theta_j \in \{\pi, \kappa, \sigma, \xi\}$:
\[
CI_{\theta_j} = \Big[ Q_{\alpha/2}\big(\{\theta_j^{*(b)}\}_{b=1}^B\big),\;
                   Q_{1-\alpha/2}\big(\{\theta_j^{*(b)}\}_{b=1}^B\big) \Big],
\]
where $Q_q(\cdot)$ denotes the empirical $q$-th quantile.
\State \Return $\hat{\theta}$ and the confidence intervals $\{CI_{\pi}, CI_{\kappa}, CI_{\sigma}, CI_{\xi}\}$.
\end{algorithmic}
\end{algorithm}

\end{document}